\documentclass[prd,aps,a4paper,twocolumn,superscriptaddress,nofootinbib]{revtex4-2}
\usepackage[a4paper,margin=2cm]{geometry}
\usepackage[caption=false]{subfig} 
\usepackage{ragged2e} 
\DeclareCaptionJustification{justified}{\justifying}
\usepackage{float}
\usepackage{amsmath}
\usepackage{graphicx}
\usepackage[colorlinks=true, allcolors=blue]{hyperref}
\usepackage{physics}
\usepackage{stmaryrd}
\usepackage{dsfont}
\usepackage{mathtools}
\usepackage[export]{adjustbox}
\usepackage[normalem]{ulem}
\usepackage{lipsum}
\usepackage{braket}         

\graphicspath{{figures/}{./}}

\newcommand{\epr}[1]{\ket{\text{\small EPR}_{#1}}}

\newcommand{\ie}{\textit{i.e.}, }
\newcommand{\eg}{\textit{e.g.}, }
\newcommand{\fepr}{F_{\text{\tiny EPR}}}

\newcommand{\size}{S_{AB}}

\begin{document}

\author{Tal Schwartzman}
\email{tal.schwartzman@cfa.harvard.edu}
\affiliation{ITAMP, Center for Astrophysics \textbar\ Harvard \& Smithsonian, Cambridge, Massachusetts 02138, USA}
\affiliation{Department of Physics, Ben-Gurion University of the Negev, David Ben Gurion Boulevard 1, Beer Sheva 84105, Israel}
\author{Antonio F.\ Rotundo}
\email{af.rotundo@gmail.com}
\affiliation{Fermioniq, Science Park 408, 1098 XH Amsterdam, The Netherlands}
\author{Raz Monsonego}
\email{razmonsonego2@gmail.com}
\author{Shira Chapman}
\email{schapman@bgu.ac.il}
\affiliation{Department of Physics, Ben-Gurion University of the Negev, David Ben Gurion Boulevard 1, Beer Sheva 84105, Israel}

\title{Multi-Boundary Many-Body Quantum Teleportation}

\begin{abstract}
Unlike standard quantum teleportation, many-body teleportation uses scrambling to transmit quantum information. In this protocol, initially localized information spreads over many degrees of freedom and is later refocused at the receiver by a simple coupling between the systems, followed by further many-body evolution. The protocol was developed from models of traversable wormholes in holography and has become a useful probe of scrambling. In particular, it can distinguish genuine scrambling from decoherence or noise when out-of-time-order correlators fail, and can reveal signatures of different scrambling mechanisms, including the distinctive behavior expected in holographic systems. Holography predicts that related protocols can transmit information between selected boundaries of multi-boundary wormhole geometries. Motivated by this setting, we study single-qubit many-body teleportation among three systems of qubits. The initial state consists of EPR pairs distributed among them, providing a simple analogue of an infinite-temperature three-boundary holographic state. We analyze the protocol with one-dimensional and all-to-all dynamics, both analytically and numerically using random circuits. We find that the third system suppresses teleportation once the spreading message reaches the region of qubits that are entangled with it. In one dimension, both the minimum coupling required for successful teleportation and the fidelity depend on the distance from the injection site to this region, a feature reminiscent of holographic causal shadows. For all-to-all dynamics, successful teleportation is instead restricted to early times and to sufficiently few qubits entangled with the third system. A third system therefore provides spatial, or subsystem, resolution of information spreading that is absent from the two-sided protocol. Our results offer a step toward many-body teleportation networks.
\end{abstract}

\maketitle

\section{Introduction}
Generic interacting quantum many-body systems can exhibit extremely complex microscopic dynamics. One manifestation of this complexity is that initially local perturbations spread over many degrees of freedom, making the information they carry increasingly inaccessible to local probes. This spreading of initially localized quantum information, known as scrambling, has become central to the study of quantum chaos and thermalization in closed systems \cite{xu2024scrambling}. Despite the complexity of the underlying dynamics, scrambling can give rise to simple coarse-grained behavior, which can be used in information-processing protocols.

One example is the so-called \emph{many-body quantum teleportation} \cite{Schuster:2021uvg, brown2023quantum, nezami2023quantum}. 
Similar to the usual quantum teleportation,  in this protocol, two parties share some entangled state, which they use as a resource to transmit quantum information. In the many-body protocol, however, the information is first scrambled through the systems by letting them evolve under their local dynamics.
At this point, the information is locally inaccessible and one would naively think that a complex decoding would be required to recover it. Surprisingly, a simple decoding protocol is instead sufficient: it is enough to briefly couple the two systems with a somewhat generic interaction and then let them freely evolve.

This counter-intuitive teleportation protocol was discovered in the context of holography \cite{gao2017traversable, maldacena2017diving}, where the shared state is dual to a two-sided wormhole that is made traversable by the coupling (see \cite{gao2019regenesis, gao2021traversable, maldacena2018eternal, bao2018traversable, freivogel2020traversable, liu2024fidelity} and the closely related works on the Hayden-Preskill protocol \cite{hayden2007black, Yoshida:2017non}). 
The protocol joins a broader body of work that studies scrambling or utilizes it as a resource, including approaches inspired by holography and black-hole information physics
\cite{Hosur:2015ylk,qi2019measuring,kobrin2024universal,vikram2026bidirectional}.  
Moreover, exploiting the connection with holography, the protocol has been proposed as a way to indirectly probe quantum gravity in the lab \cite{nezami2023quantum, brown2023quantum}, with
first experimental explorations including \cite{Landsman:2018jpm, blok2021quantum, jafferis2022traversable}.

The microscopic mechanism behind the protocol was investigated
in \cite{Schuster:2021uvg, brown2023quantum, nezami2023quantum}, where
it was shown that its success depends on how operators spread under
the system's time evolution. Thus, unlike in the conventional
teleportation protocol, successful teleportation is not guaranteed by
the protocol itself, and indicates nontrivial dynamical properties of the
system. For certain parameter regimes of the protocol, scrambling systems generically admit
successful teleportation.
On the other hand, the protocol's specific information transfer capabilities and quantitative behavior can differ between systems with distinct scrambling dynamics, thereby providing a novel probe of scrambling and operator size distributions \cite{Yoshida:2018vly, Schuster:2021uvg, Landsman:2018jpm}.

One mechanism that enables teleportation and is generically present in scrambling systems is the \textit{peaked-size mechanism}. It is based on the idea that when local operators encoding the message are sufficiently scrambled over a large system, their (relative) size distributions become sufficiently narrow, thereby explaining how the coupling uniformly affects the message and induces teleportation. Another interesting mechanism, known as  \textit{size winding}, is thought to be responsible for the distinctive performance of the protocol when holographic systems are used as a resource \cite{Schuster:2021uvg, brown2023quantum, nezami2023quantum, zhou2024size, perugu2025krylov}.

In this article, we generalize the many-body quantum teleportation protocol to three parties. Our motivations are as follows. First, the additional party enhances the protocol's capabilities as a probe of scrambling. While the teleportation fidelity in the two-party case already contains information about the evolving operator-size distribution, in the three-party setting it also becomes sensitive to which sites support the time-evolved message operators, thereby providing spatial or subsystem resolution of information spreading. Furthermore, considering more than two parties elevates the two-party setup to a quantum network, where one can use the protocol to communicate between the different nodes.

Finally, in holography, a number of papers have studied multi-boundary wormholes and how to make them traversable \cite{balasubramanian2014multiboundary, marolf2015hot, al2021traversability,emparan2021multi}.
A qualitative difference between these geometries and the usual two-boundary wormhole studied in \cite{gao2017traversable} is the presence of \emph{causal shadows} \cite{headrick2014causality}.
These are regions of spacetime that are causally disconnected from all boundaries and are expected to make traversing the wormhole more difficult in the following sense. 
In the high-temperature limit, there are certain regions within the boundaries whose local entanglement structure resembles that of the two-boundary wormhole state
\cite{al2021traversability}. 
If we restrict the teleportation protocol to these regions, we expect it to succeed with good fidelity. On the other hand, it becomes increasingly difficult to teleport the message as we move away from them. Here we investigate whether a simpler circuit toy-model can reproduce this behavior and shed light on its microscopic origin.

Motivated by the above considerations, we study a three-party many-body teleportation protocol in which each party holds a qubit system.\footnote{See Sec.~\ref{sec:qudit_generalization} for a generalization involving qudits.} We focus on a single qubit teleportation between two of the three parties. The qubit systems will have either one-dimensional local dynamics or zero-dimensional all-to-all dynamics. 
Similarly to the two-party protocol, the protocol involves the coupling of these two systems. The initial state is built from EPR pairs distributed among all three parties, providing a simple analogue of the infinite-temperature three-boundary holographic state.  We analyze the protocol's outcome analytically and numerically. Numerical simulations are performed for circuits composed of two-qubit random gates, either in a brickwork ($1D$) or random all-to-all ($0D$) architecture.

In one dimension, both the minimum coupling required for
successful teleportation and the fidelity depend on the distance from the injection site to the sites entangled with the third party. For all-to-all dynamics, successful
teleportation is instead restricted to early times and to sufficiently few qubits entangled with the third system. 
The resulting behavior reflects a competition between the peaked-size mechanism, which improves the teleportation fidelity as the message scrambles, and the spreading of the message into degrees of freedom entangled with the third party, which suppresses it.  Our results reproduce the phenomenology expected from holographic three-boundary traversable wormholes, and show how this setup indeed provides additional spatial or subsystem resolution of the spreading operator.

This paper is organized as follows. Section \ref{sec:prot} presents the three-party protocol and its holographic motivation, introduces the expressions needed for the teleportation fidelity, and details the effect of the coupling. Section \ref{sec:fid} derives a lower bound on the fidelity and identifies the competition between peaked-size teleportation and the spread of the message into degrees of freedom entangled with the third party. Section \ref{sec:radncir} studies this competition analytically and numerically in one-dimensional brickwork (subsection \ref{sec:radncir1d}) and all-to-all (subsection \ref{sec:radncir0d}) random circuits. Section \ref{sec:qudit_generalization}  analyzes the effect of increasing the local dimension of the resource systems while keeping a single-qubit message. Section \ref{sec:concl}  summarizes our results and outlines directions for future work.

\section{The protocol}\label{sec:prot}
\begin{figure}
    \centering
    \includegraphics[width=0.5\linewidth]{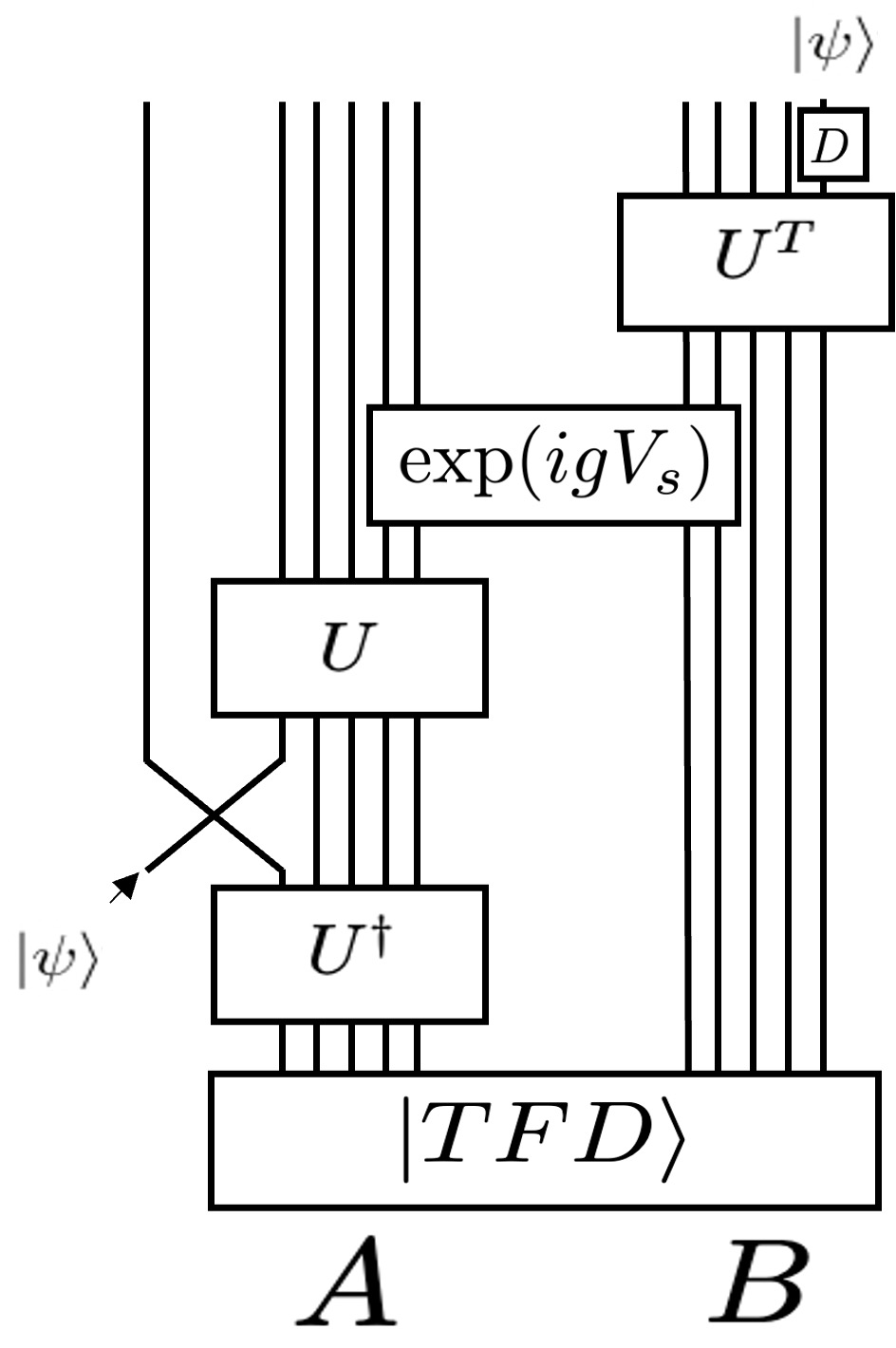}
    \caption{A diagram of the two-sided protocol. The lines represent degrees of freedom, with the circuit evolving upwards.}
    \label{fig:twosided}
\end{figure}

Before introducing the protocol considered in this work, we briefly describe the standard wormhole teleportation, which can be seen in Fig.\ \ref{fig:twosided}.
At time $t=0$, two parties, $A$ and $B$, share a specific entangled state, called the thermofield double (TFD) state,
\begin{equation}
\ket{\text{TFD}} = \frac{1}{\sqrt{Z_\beta}}\sum_{n}e^{-\beta E_n/2}\ket{n}\otimes \ket{n^*},
\end{equation}
whose dynamics are generated by a Hamiltonian $H\otimes \textbf{1}+\textbf{1} \otimes H^T$.
Above, $\ket{n}$ is an eigenstate of $H$ with energy $E_n$, $\ket{n^*}$ is its complex conjugate, and $Z_\beta=\sum e^{-\beta E_n}$ is the canonical partition function.\footnote{The complex conjugation and transposition are defined with respect to a certain basis of the Hilbert space. Here, we take this basis to be the computational basis, such that at infinite temperature, the state is a product of EPR pairs.} 
The TFD state is such that the reduced density matrix on both $A$ and $B$ is thermal at inverse temperature $\beta$.
In the context of holography, this state is dual to a two-sided eternal black hole if the temperature is above the Hawking-Page phase transition \cite{ maldacena2003eternal, balasubramanian2014multiboundary}.

The wormhole teleportation protocol works as follows.  
First, we inject the message into the $A$ system at time $-t$, and evolve the state to $t=0$.
Since the dynamics are scrambling, at this point the message is effectively lost to both $A$ and $B$.
We would naively expect that a complex decoding unitary should be applied to recover it.
Surprisingly, as first discovered in \cite{gao2017traversable}, it is enough for $A$ and $B$ to briefly couple their systems with a generic unitary - \ie simple and independent of the message sent - and then let them evolve freely to time $t$ to see the message refocusing on B’s side. We discuss the precise form and role of the coupling $V_s$ and the local decoding unitary $D$ in Sec. \ref{subsec:Succ}.

It is natural to ask how this protocol can be generalized to the multi-party case. 
Namely, consider now three parties, $A$, $B$ and $C$, initially sharing some entangled state $\ket{\Psi}$.
The goal of the protocol is to transmit a number of qubits from one party to another, say A to B. 
Several works have studied this question in holography 
\cite{balasubramanian2014multiboundary, marolf2015hot, al2021traversability,emparan2021multi}, but less is known in generic quantum mechanical models. 

The first problem one faces is which state to use to replace the TFD.
Generalizations of the TFD state to multiple parties were first considered in the context of holography \cite{balasubramanian2014multiboundary, marolf2015hot}, and later for tensor network models \cite{peach2017tensor} and critical spin chains \cite{zou2022multiboundary}. 

\begin{figure}
    \centering
    \includegraphics[width=0.5\linewidth]{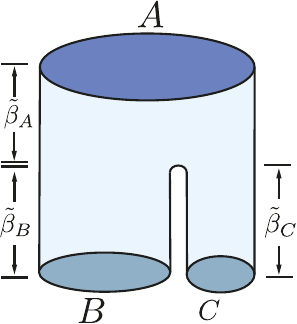}
    \caption{The pair of pants geometry for the path integral representation of the wave function associated with the three-party generalization of the TFD state.}
    \label{fig:PathInt}
\end{figure}

As a guiding example, \cite{zou2022multiboundary} presented the case of three systems $A,B$ and $C$ where $\text{dim}\,{A} = \text{dim}\, {B}\times \text{dim}\, {C}$.
In this case, the construction of the three-party generalized TFD state can be visualized in terms of a pair of pants path integral representation (see Fig. \ref{fig:PathInt}). 
The state takes the form
\begin{equation}
\begin{split}
\label{eq:3_wormhole}
&\ket{\Sigma_3}  = \frac{1}{\sqrt{Z}} e^{-\tilde \beta_A H_A}\otimes e^{-\tilde \beta_B H_B} \otimes e^{-\tilde \beta_C H_C}\ket{\Omega}
    \\
    & =\sum_{ijk}C_{ijk}e^{-\tilde{\beta}_A E_i^A}e^{-\tilde{\beta}_B E_j^B}e^{-\tilde{\beta}_C E_k^C}\ket{E_i^A E_j^B E_k^C}.
\end{split}
\end{equation}
Here $H_A$, $H_B$, and $H_C$ are the Hamiltonians of the three systems, and $\ket{\Omega}$ is a maximally entangled state between system $A$ and the combined system $BC$. For example, in a spin chain one may write
\begin{equation}
\ket{\Omega}=\sum_{i,\tilde i}\ket{i\tilde i}_A\ket{i}_B\ket{\tilde i}_C,
\end{equation}
where $i$ and $\tilde i$ label states in the computational basis. The second line of Eq.~\eqref{eq:3_wormhole} follows by expanding the state in the energy eigenbasis $\ket{E_i^A E_j^B E_k^C}$. The parameters $\tilde\beta_A$, $\tilde\beta_B$, and $\tilde\beta_C$ are Euclidean evolution times, which determine the lengths of the different segments of the pair of pants geometry, see Fig.~\ref{fig:PathInt}.
In the limit $\tilde\beta_i\to 0$, the path integral locally identifies field configurations in the different systems. For critical systems, the coefficients $C_{ijk}$ are fixed by the conformal data. In holographic systems, such as those studied in Refs.~\cite{balasubramanian2014multiboundary, marolf2015hot}, the dual description gives this state a geometric interpretation as a spacetime with three boundaries, where for suitable ranges of the parameters, this geometry is a fully connected three-boundary wormhole.

In this work, we would like to consider closely related configurations in which the three systems play a symmetric role. The three systems will be chains of qubits. 
Let $2n$ be the number of qubits held by each party, $I\in \{A, B, C\}$.
We denote the qubits of $I$ as $I_i$ with $i=1, \dots, 2n$.
It is convenient to place them on a circle, such that $I_{i}=I_{2n+i}$.
We focus on the infinite-temperature setting and leave finite-temperature corrections for future work. With the above assumptions, the state is given by a set of EPR pairs shared among the different parties, where each pair of parties shares $n$ EPR pairs. 
The initial state of the system is 
\begin{equation}\label{PsiDef}
    \ket{\Psi}=\epr{AB}\otimes \epr{BC}\otimes \epr{CA},
\end{equation}
where 
\begin{align}
\ket{\mathrm{EPR}_{AB}}
&:=
\bigotimes_{k=1}^{n}
\ket{\mathrm{EPR}_{A_k B_k}},
\\
\ket{\mathrm{EPR}_{BC}}
&:=
\bigotimes_{k=1}^{n}
\ket{\mathrm{EPR}_{B_{n+k} C_k}},
\\
\ket{\mathrm{EPR}_{CA}}
&:=
\bigotimes_{k=1}^{n}
\ket{\mathrm{EPR}_{C_{n+k} A_{n+k}}},
\label{EPRdef}
\end{align}
and
\begin{equation}
\epr{I_a J_b}=\frac{1}{\sqrt{2}}\left(\ket{0}_{I_a}\otimes\ket{0}_{J_b}+\ket{1}_{I_a}\otimes\ket{1}_{J_b}\right).
\end{equation}

This state is simple enough that we will manage to do some calculations semi-analytically, but still retains some interesting geometrical structure and can be considered a toy model capturing certain aspects of multi-boundary holographic wormholes \cite{marolf2015hot, peach2017tensor, zou2022multiboundary}.

In the infinite-temperature limit, the initial state has no genuine tripartite entanglement and can instead be viewed as three infinite-temperature two-sided TFDs. Nevertheless, the imposed dynamics are not equivalent to the dynamics of three independent infinite-temperature two-sided TFDs, since they do not factorize accordingly.

The protocol then follows the same steps as in the two-party case \cite{Schuster:2021uvg}.
To simplify the calculation, rather than considering the teleportation of a specific state, we will consider the entanglement swap.
In other words, Alice holds an EPR pair that she wants to share with Bob.
To do this, at time $-t$, she swaps one half of her EPR pair, denoted by $R_1$, with one of the qubits of system $A$, denoted by $A_r$, and then evolves the state to $t=0$ by applying $U_t$. 
A coupling between $A$ and $B$ is performed using a unitary $V$, and then the state of system $B$ is evolved forward to time $t$ using $U_t^T$, the transpose of the evolution operator of system $A$. 
A final decoding unitary $D$ is applied to the qubit $B_r$ of system $B$, the one which in $\ket{\Psi}$ is paired with $A_r$. 
If the protocol worked correctly, $B_r$ should now be entangled with the half EPR pair left with Alice  \ie with $R_0$. For notational convenience, from now on we will often omit the time subscript on the evolution operator and write $U$ instead of $U_t$. 
In the diagrammatic notation of \cite{Yoshida:2017non, Schuster:2021uvg}, the protocol is
\begin{align}\label{finalstate}
\begin{gathered} 
\centering
\includegraphics[width=0.4\textwidth,valign=c]{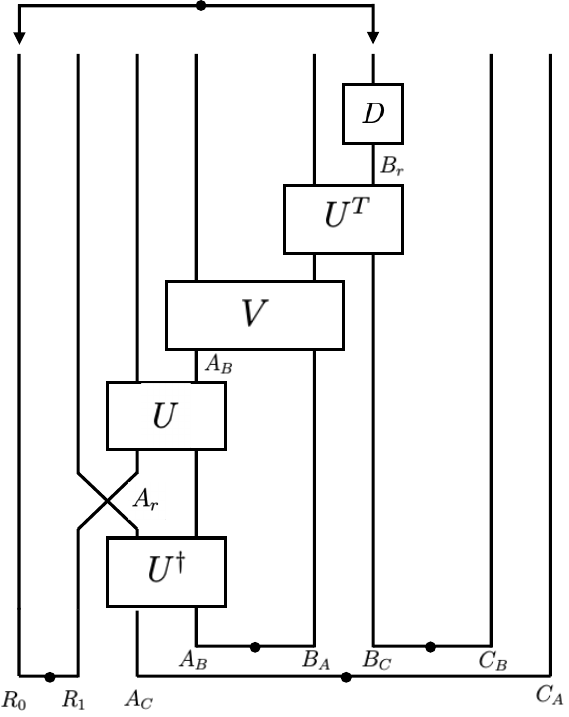}.
\end{gathered}
\end{align}
Here time flows upwards, the legs $I_J$ represent the qubits of $I$ entangled with $J$, $R_{0}$, $R_1$, $A_r$ and $B_r$ each represent a single qubit, and a dot represents a normalization factor according to the number of EPR pairs the leg represents, \ie $2^{-1/2}$ for the leg connecting $R_0$ and $R_1$, and $2^{-n/2}$ for the rest. Let us emphasize that the output legs of the unitaries in the diagram do not necessarily represent the same ordering of qubits as the input legs. For example, unlike the input legs of the first unitary $U^\dagger$, the left output leg represents a single qubit at site $r$, and the right leg represents the rest of the qubits.
From now on, we will often resort to this type of diagrammatic notation to represent equations.
Notice that upon tracing away system $C$, the protocol reduces to the two-boundary case but with an initial mixed state for the two remaining boundaries.\footnote{As long as system $C$ is not involved in the protocol, one can alternatively view the $A_C$ and $B_C$ sites as a specific type of environment to which the information leaks in the two-sided protocol. See \cite{zhou2024environment} for further results on the effect of environment in the many-body quantum teleportation with a different setup.} 

The coupling between the two sides will be of the form
\begin{equation}\label{eq:V}
    V \equiv e^{i g V_s},
\end{equation} 
where $V_s$ couples the pairs of sites that were maximally entangled in the initial state. We describe it in detail later on.  

\subsection{Expectations from holography}

\begin{figure}
    \centering
    \includegraphics[width=0.7\linewidth, trim=0.8 0 0.8 0.8cm,
  clip]{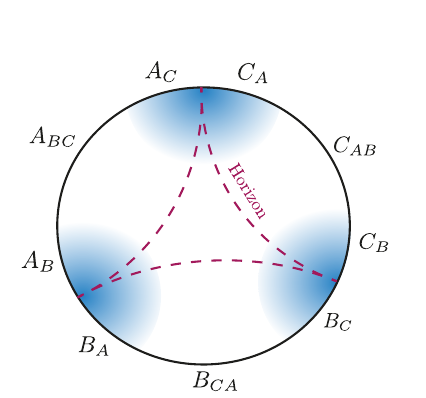}
    \caption{Schematic representation of the $t=0$ slice of a three-boundary wormhole. The dashed lines are the black hole horizons. 
    In the boundary regions adjacent to the blue bulk regions, the entanglement is approximately bipartite. The bulk region bounded by the red dashed lines is the causal shadow.}
    \label{fig:causal shadow}
\end{figure}

One qualitative difference between the two-sided and the three-sided case is that in the latter, there exists a region between the horizons that is causally disconnected from all boundaries, a so-called \emph{causal shadow}.
As explained in \cite{al2021traversability}, the geometry of the system on the $t=0$ spatial slice can be schematically represented as in Fig.\ \ref{fig:causal shadow}.
The three asymptotic regions are separated by three horizons, and the region in between them is precisely the causal shadow. 
In the boundary regions adjacent to the blue bulk regions in Fig.\ \ref{fig:causal shadow}, the entanglement is bipartite within an error exponentially small in the temperature, and the state is well approximated by the TFD. 
Correspondingly, the causal shadow becomes exponentially thin in these regions.
We denote by $I_J$ the region of $I$ which is approximately only entangled with $J$.
These regions are separated by regions where tripartite entanglement cannot be neglected, \eg $A_{BC}$ in Fig.\ \ref{fig:causal shadow}. 
As shown in \cite{marolf2015hot}, in the limit of large temperatures $\beta_I\rightarrow 0$ with fixed ratios $\beta_I/\beta_J$, the size of these regions goes to zero, and the entanglement of the overall state becomes bipartite within an error exponentially small in the temperature.

\begin{figure}
    \centering
    \includegraphics[width=0.8\linewidth]{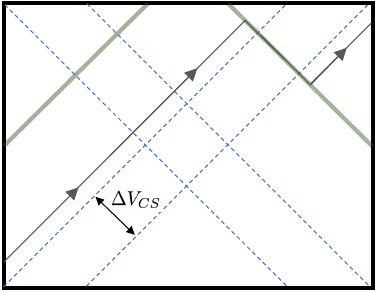}
    \caption{The Penrose diagram of a black hole spacetime having a causal shadow with a particle that traverses it. This is a schematic representation of the causal structure of a section that contains two asymptotic regions in
    the three-boundary wormhole geometry \cite{al2021traversability}. The particle has a spacelike region to traverse that depends on the width of the causal shadow in this section. In the traversable wormhole protocol, 
    the coupling induces negative averaged null
    energy and a corresponding null shift. This null shift should be large enough for the particle to cross the
    null-coordinate gap of the causal shadow.}
    \label{fig:PenroseCS}
\end{figure}

To make the multi-boundary wormhole traversable, Ref.~\cite{al2021traversability} introduces a coupling between two of the boundary systems, analogous to the double-trace coupling in the two-boundary construction \cite{gao2017traversable}. In the bulk, this coupling generates a negative-energy perturbation that shifts the trajectory of an excitation by an amount $\Delta V$ along a null direction. For the excitation to reach the receiving boundary, this shift must overcome the separation $\Delta V_{\mathrm{CS}}$ produced by the causal shadow, as illustrated in Fig.~\ref{fig:PenroseCS}. Since $\Delta V$ increases with the coupling strength, a trajectory that crosses a wider portion of the causal shadow requires a stronger coupling. The required coupling therefore depends on the angular position of the insertion point (cf. Fig.~\ref{fig:causal shadow}) and increases as the insertion point moves away from the region predominantly entangled with the receiving boundary and toward the region entangled with the third boundary. This is the qualitative behavior that we reproduce and explain in our model. We stress, however, that our model is far from having a semiclassical gravitational dual. Nevertheless, the microscopic mechanism responsible for this behavior within our protocol may also be relevant in holographic systems.

\subsection{Success probability}\label{subsec:Succ}
The success probability (fidelity)  of the protocol is given by the expectation value of the projector on the EPR state between $R_0$ and $B_r$,
\begin{align}
\fepr=\includegraphics[height=11cm,valign=c]{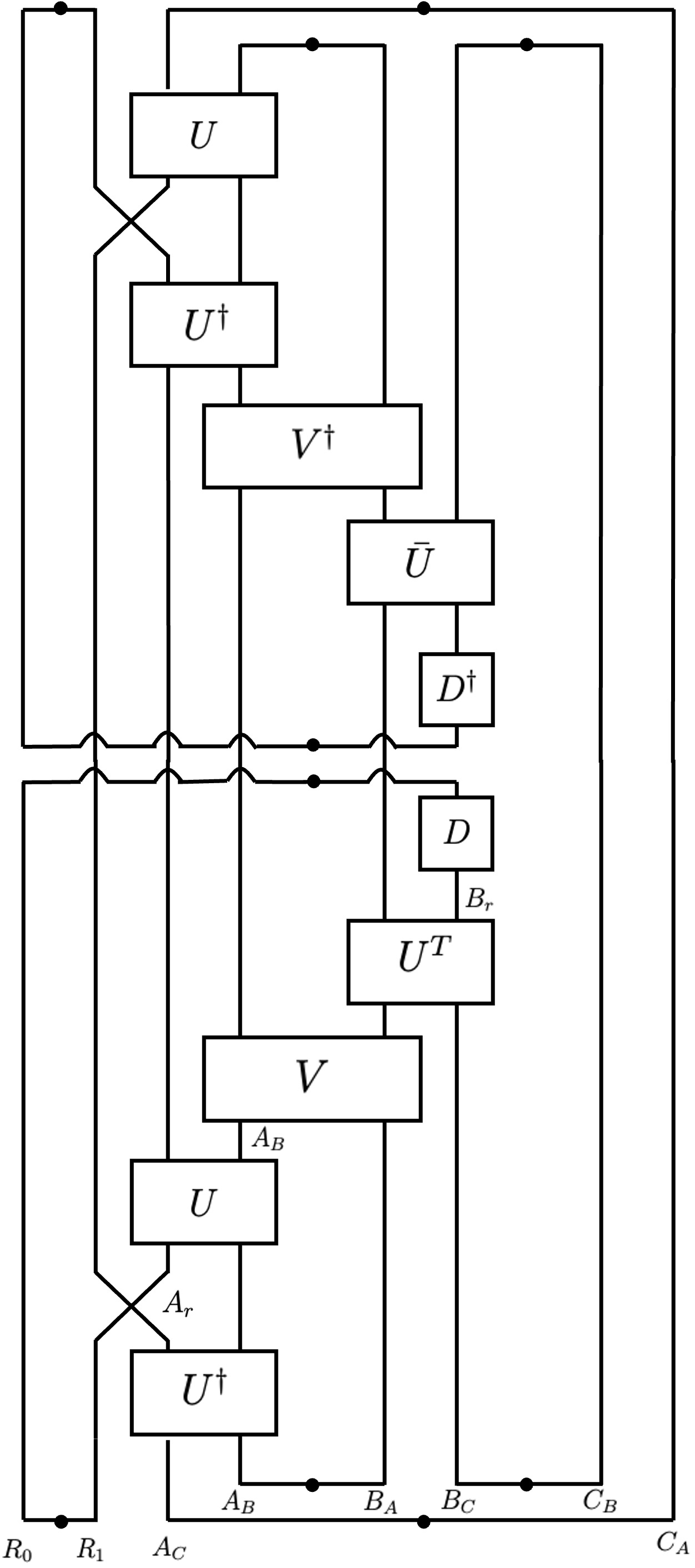},
\end{align}
where the bar denotes complex conjugation.  We can simplify the circuit above by expressing the swap operator as $\text{SWAP} = \frac{1}{2}\sum_\mu \sigma_\mu \otimes \sigma_\mu$, where $\sigma_\mu$ is one of the Pauli operators, including the identity.
We find
\begin{equation}  
\fepr= \sum_{\mu \nu}\frac{1}{4}
\begin{gathered}
\includegraphics[height=11cm,valign=c]{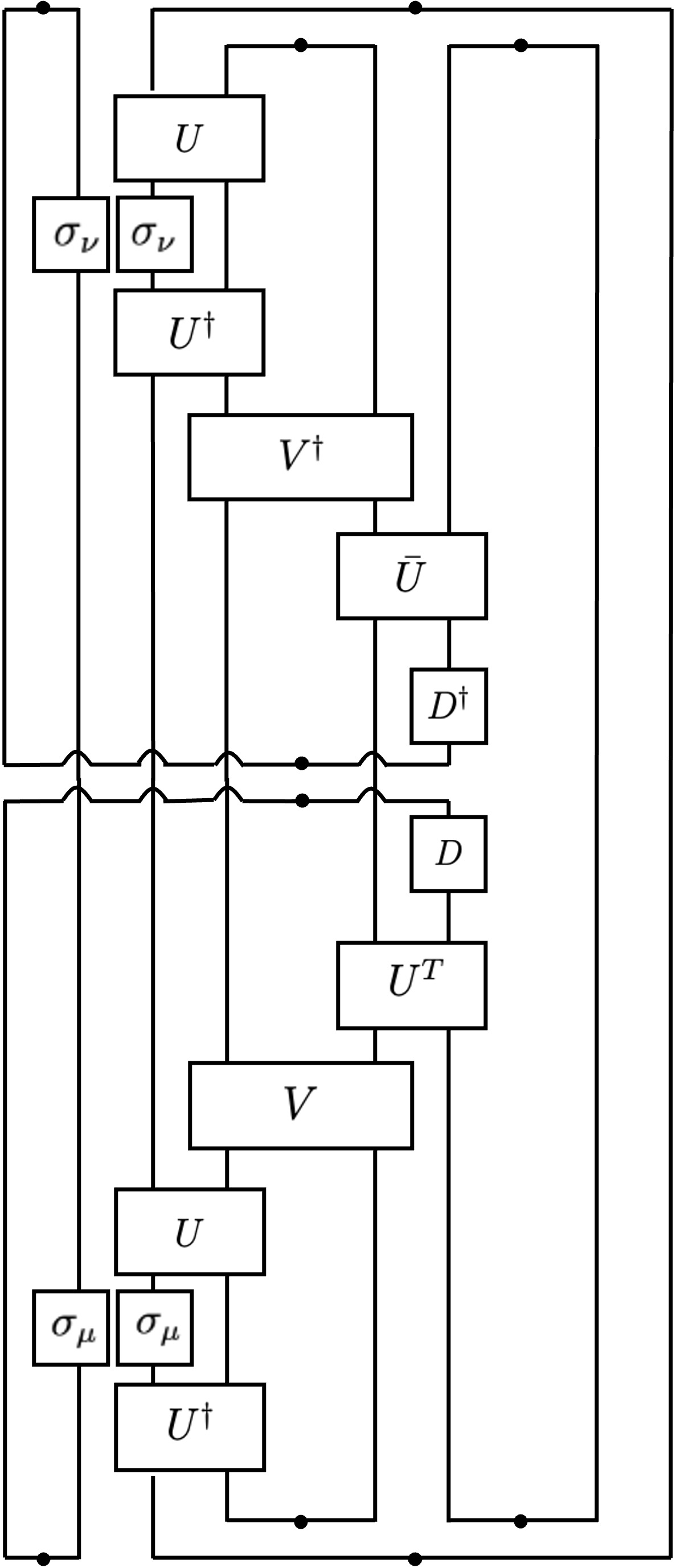} 
\end{gathered}\;.
\end{equation}
Reshaping the line connecting $D$ and $D^\dagger$, and taking out the factors $1/\sqrt{2}$ associated with the dots, we find 
\begin{equation}\label{eq:FEPR3U}
\fepr= \sum_{\mu \nu}\frac{1}{2^4}
\begin{gathered}
\includegraphics[height=11cm,valign=c]{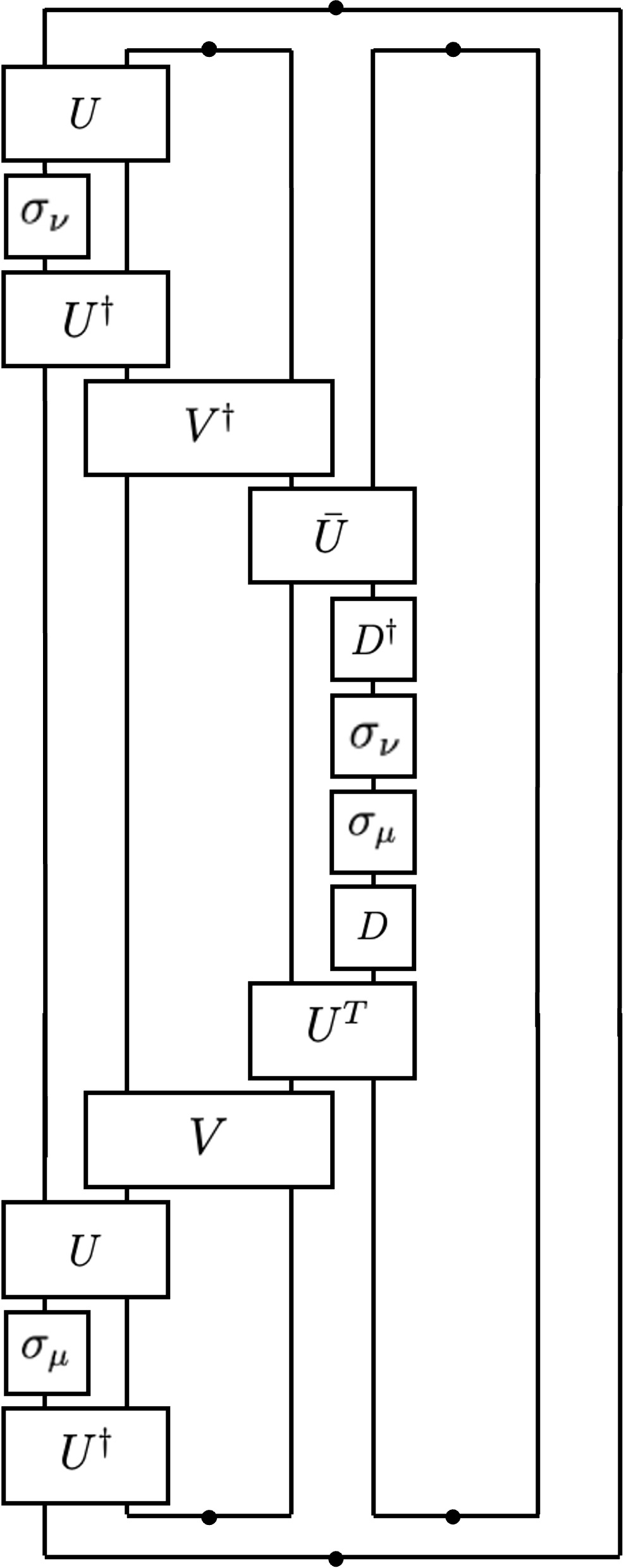}
\end{gathered}\;.
\end{equation}
Finally, inserting the identity $DU^T \bar{U}D^\dagger$ between $\sigma_\mu$ and $\sigma_\nu$ in the middle of the diagram, we arrive at 
\begin{equation}
\label{fidelityExact}
    2^4\fepr  = \sum_{\mu\nu}
    \bra{\Psi} \sigma^{A_r}_{\nu,t} V^\dagger \tilde \sigma^{B_r}_{\nu,t}  \tilde \sigma^{B_r}_{\mu,t} V \sigma^{A_r}_{\mu,t}\ket{\Psi}.
\end{equation}
Here we have defined
\begin{equation}\label{evolvedPauli}
    \begin{split}
    & \sigma^{A_r}_{\mu,t}  \equiv \sigma^{A_r}_{\mu}(-t)=U_t\sigma^{A_r}_\mu U_t^\dagger,
    \\
    & \tilde \sigma^{B_r}_{\mu,t} \equiv \bar U_t D^\dagger \sigma^{B_r}_\mu D U_t^T,
    \end{split}
\end{equation}
 where the superscript denotes the Hilbert space on which the operators act at $t=0$. 
In other words, at $t=0$, $\sigma^{A_r}_{\mu,t}$  has support only on the qubit $A_r$, while at later times it spreads to other qubits according to the evolution operator $U_t$.
Notice that the operators only spread to other qubits within their local system, \ie within system $A$ for $\sigma^{A_r}_{\mu,t}$, and system $B$ for $\tilde \sigma^{B_r}_{\mu,t}$. 

In what follows, it will be useful to define the following quantity for system $B$, analogous to  $\sigma^{A_r}_{\mu,t}$: 
\begin{equation}
    \sigma^{B_r}_{\mu,t}  \equiv U_t \sigma^{B_r}_\mu U_t^\dagger.
\end{equation}
Being explicit about the entire Hilbert space, these operators can be expressed as 
\begin{equation} \label{SigBcopySigA}
\begin{split}
    &\sigma^{A_r}_{\mu,t} = U_t\sigma^{r}_\mu U_t^\dagger \otimes \boldsymbol{1}_B \otimes \boldsymbol{1}_C,
    \\
    & \sigma^{B_r}_{\mu,t} = \boldsymbol{1}_A \otimes  U_t \sigma^{r}_\mu U_t^\dagger \otimes \boldsymbol{1}_C.
\end{split}
\end{equation}
We emphasize that $\sigma^{B_r}_{\mu,t}$ is just a copy of $\sigma^{A_r}_{\mu,t}$ on the Hilbert space of system $B$.

It will also be useful to  denote a Pauli string that acts on $2n$ sites by $\sigma_{\bar \mu}$, \ie
\begin{equation}\label{PauliStringGen}
\sigma_{\bar{\mu}} = \bigotimes_{k=1}^{2n}\sigma_{\mu_k},
\end{equation}
with $\mu_k\in \{0,1,2,3\}$, \ie  $\sigma_{\mu_k} = (\boldsymbol{1}, \sigma_x, \sigma_y, \sigma_z)$.

We now move on to specify the coupling $V = e^{i g V_s}$ between the $A$ and $B$ systems. 
As argued in \cite{Schuster:2021uvg}, a successful protocol can be achieved with a fairly generic interaction that consists of a sum over couplings between the many pairs that are initially maximally entangled. This is because the scrambling dynamics of the system cause generic couplings of this type to have approximately the same effect at late times. This common effect is captured by an interaction proportional to the size operator, which we therefore use below: 
\begin{equation}\label{interactionHam}
    V_s =  \frac{1}{n} \sum_{i=1}^{n}\left(P_{AB,i} - \boldsymbol{1}\right),
\end{equation}
where  $P_{AB,i}=\ketbra{\text{\small EPR}_{A_{i}, B_{i}}}$ is the EPR projector between qubits $A_{i}$ and $B_{i}$. We note that the $\boldsymbol{1}$ term in $V_s$ generates a constant global phase when applied to any state in the Hilbert space and was added for convenience. Taking $V_s \to V_s +\boldsymbol{1}$ will therefore not alter any physical predictions. To explain the effect of the coupling $V$, let us first note that the eigenstates of $V_s$ are of the form $\sigma_{\bar{\mu}}^{A_B} \otimes \boldsymbol{1}_{B_A} \epr{AB}$, where $\sigma_{\bar{\mu}}^{A_B}$ is a Pauli string acting on the sites of $A$ that are paired with $B$, namely
\begin{equation} \label{PauliString}
\sigma^{A_B}_{\bar{\mu}} = \bigotimes_{k=1}^{n}\sigma_{\mu_k}.
\end{equation}
One can check that $\sigma_{\bar{\mu}}^{A_B}\otimes \boldsymbol{1}_{B_A}\epr{AB}$ is an eigenstate of $P_{AB,i}$ with eigenvalue $1$ if $\sigma_{\bar{\mu}}^{A_B}$ acts as the identity on site $i$, and eigenvalue $0$ otherwise. 
Therefore,
\begin{equation} \label{VsAction}
V_s\cdot\sigma_{\bar{\mu}}^{A_B} \epr{AB} = -\frac{\size(\sigma_{\bar{\mu}}^{A_B})}{n}\,\sigma_{\bar{\mu}}^{A_B}  \epr{AB},
\end{equation}
where $\size(\sigma_{\bar{\mu}})$ measures the size on $A_B$ of a Pauli string $\sigma_{\bar{\mu}}$, \ie the number of nonidentity operators in the Pauli string $\sigma_{\bar{\mu}}$ within the first $n$ sites of $A$, which are those that are paired with $B$ in $\ket{\Psi}$.
Note that for a Pauli string with support on the entire system $A$,  the state $\sigma_{\bar{\mu}} \otimes \boldsymbol{1}_B \otimes \boldsymbol{1}_C \ket{\Psi}$ is an eigenstate of $V_s$, with an eigenvalue of $-\size(\sigma_{\bar{\mu}})/n$. 

Interactions like $V$ are a key ingredient in the mechanism of many-body quantum teleportation, as the size distribution of operators evolves predictably under strongly coupled dynamics.

Next, we need to specify the decoding unitary $D$.
Following the two-boundary protocol \cite{Schuster:2021uvg}, we take $D$ to be $\sigma_2$. 
Its effect on the Pauli matrices, $\sigma_2 \sigma_\mu \sigma_2$ is that it flips the sign of $\sigma_1$ and $\sigma_3$, and does nothing to the rest, \ie 
\begin{equation}
D^\dagger \sigma_\mu D=(-1)^{1+\delta_{\mu,0}}\sigma_\mu^T.
\end{equation}
It follows that 
\begin{equation}\label{tilsigBr}
\begin{split}
\tilde{\sigma}_{\mu,t}^{B_r} & =(-1)^{1+\delta_{\mu,0}}(\sigma_{\mu,t}^{B_r})^T 
\\
& = (-1)^{1+\delta_{\mu,0}}\boldsymbol{1}_A \otimes (U_t\sigma^{r}_\mu U_t^\dagger)^T\otimes \boldsymbol{1}_C.
\end{split}
\end{equation}

\section{Fidelity estimation}\label{sec:fid}
We shall now turn to derive a lower bound for the exact fidelity, given in \eqref{fidelityExact}. 
As we will soon explain, the lower bound is given by $\fepr \geq F_{\text{pure}}$, where
\begin{equation}\label{Fpure}
    F_{\text{pure}} = \abs{\sum_\nu \frac{1}{2^2} \includegraphics[height=5.9cm,valign=c]{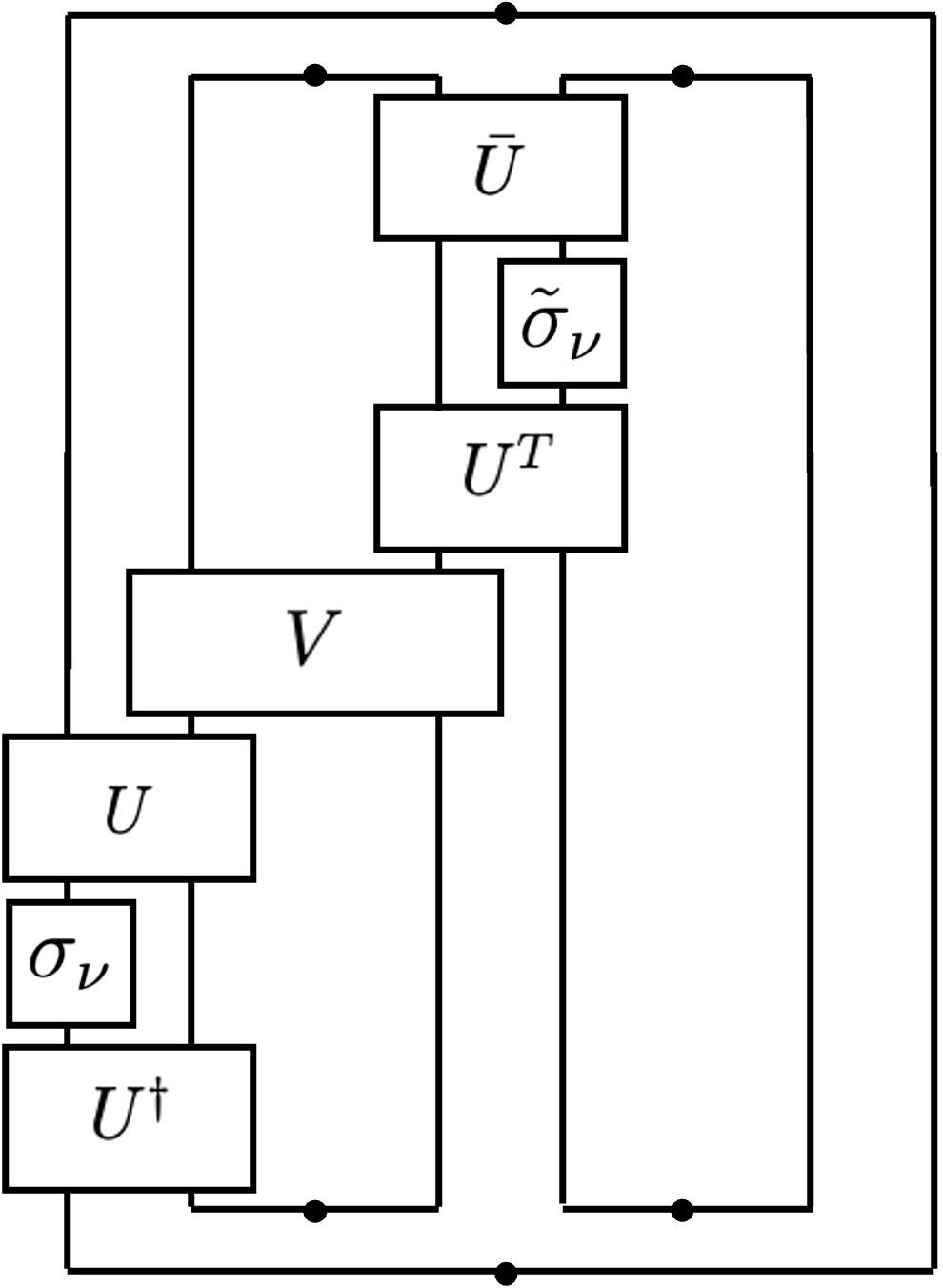} \, }^{\textstyle 2}.
\end{equation}
$F_{\text{pure}}$ has a physical interpretation of its own. When computing the fidelity in \eqref{fidelityExact}, we trace over the degrees of freedom that are not $R_0$ and $B_r$, and therefore the protocol can succeed regardless of their state. On the other hand, one might be interested in a protocol in which the final state looks exactly as if the original information (up to a decoding transformation $D$) was injected to site $B_r$ at time $t$, instead of site $A_r$ at time $-t$. The fidelity of such a protocol is given by \eqref{Fpure}.
$F_{\text{pure}}$ will provide a qualitative intuition for the fidelity's general behavior.  

Notice that \eqref{fidelityExact} can be expressed as $2^4 \fepr = \braket{V|V}$, where $\ket{V} = \sum_{\mu} \tilde \sigma^{B_r}_{\mu,t} V \sigma^{A_r}_{\mu,t}\ket{\Psi}$ is an unnormalized state. The norm of  $\ket{V}$ is greater than or equal to the norm of its projection on the normalized state $\ket{\Psi}$, \ie $\braket{V|V}\geq \braket{V|\Psi}\braket{\Psi| V}$ and therefore,
\begin{equation}
\begin{split}
    2^4 \fepr &\geq \Bigl\vert\sum_\mu \bra{\Psi}\tilde \sigma^{B_r}_{\mu,t} V \sigma^{A_r}_{\mu,t}\ket{\Psi}\Bigr\vert^2 \\
    &= \Bigl\vert\sum_\mu C_{\mu}\Bigr\vert^2 = 2^4 F_{\text{pure}},
\end{split}
\end{equation} 
where we have defined $C_{\mu} = \bra{\Psi}\tilde \sigma^{B_r}_{\mu,t} V \sigma^{A_r}_{\mu,t}\ket{\Psi}$.

In general, $\sigma^{A_r}_{\mu,t}$ and $\tilde{\sigma}_{\mu,t}^{B_r}$ can be expanded in terms of Pauli strings in system $A$ and $B$ respectively as defined in \eqref{PauliStringGen}, \ie 
\begin{equation}\label{opExp}
\begin{split}
\sigma^{A_r}_{\mu,t} & = \sum_{\bar \nu} c_{\mu,\bar \nu}(t) \sigma_{\bar{\nu}} \otimes \boldsymbol{1}_B \otimes \boldsymbol{1}_C,
\\
\tilde{\sigma}_{\mu,t}^{B_r} &  = (-1)^{1+\delta_{\mu,0}}\sum_{\bar \nu} c_{\mu,\bar \nu}(t)\boldsymbol{1}_A \otimes \sigma_{\bar{\nu}}^T\otimes \boldsymbol{1}_C,
\end{split}
\end{equation}
where we have used \eqref{tilsigBr}. By unitarity, the different $c_{\mu,\bar \nu}(t)$  coefficients are real, and their squares form a probability distribution,
\begin{equation}\label{eq:c_iNormalization}
    \sum_{\bar \nu} c_{\mu,\bar \nu}(t)^2 = 1.
\end{equation} 

Computing $C_{\mu}$ gives 
\begin{equation}
\begin{split}
    & C_{i} =
    \\
    & -\sum_{\bar \mu,\bar \nu}e^{-i g \size(\sigma_{\bar \mu})/n} c_{i,\bar \mu} c_{i,\bar \nu} \bra{\Psi} \sigma_{\bar \mu} \otimes \sigma_{\bar \nu}^T \otimes \boldsymbol{1}_C \ket{\Psi},
    \\
    & C_{0} = 1,
\end{split}
\end{equation}
where the time dependence of coefficients $c_{i,\bar \mu} = c_{i,\bar \mu}(t)$ is implicit. Notice that this is the first place in which the choice of the initial state $\ket{\Psi}$ in \eqref{PsiDef} has played a role, allowing the coupling to act in the simple way given by \eqref{VsAction}.
Let us now focus on the $\bra{\Psi} \sigma_{\bar \mu} \otimes \sigma_{\bar \nu}^T \otimes \boldsymbol{1}_C \ket{\Psi}$ term.
We can diagrammatically express it as, 
\begin{equation} \label{expectToTrace}
\begin{split}
\bra{\Psi} & \sigma_{\bar \mu} \otimes  \sigma_{\bar \nu}^T \otimes \boldsymbol{1}_C  \ket{\Psi}      \\
    & = \includegraphics[height=2.3cm,valign=c]{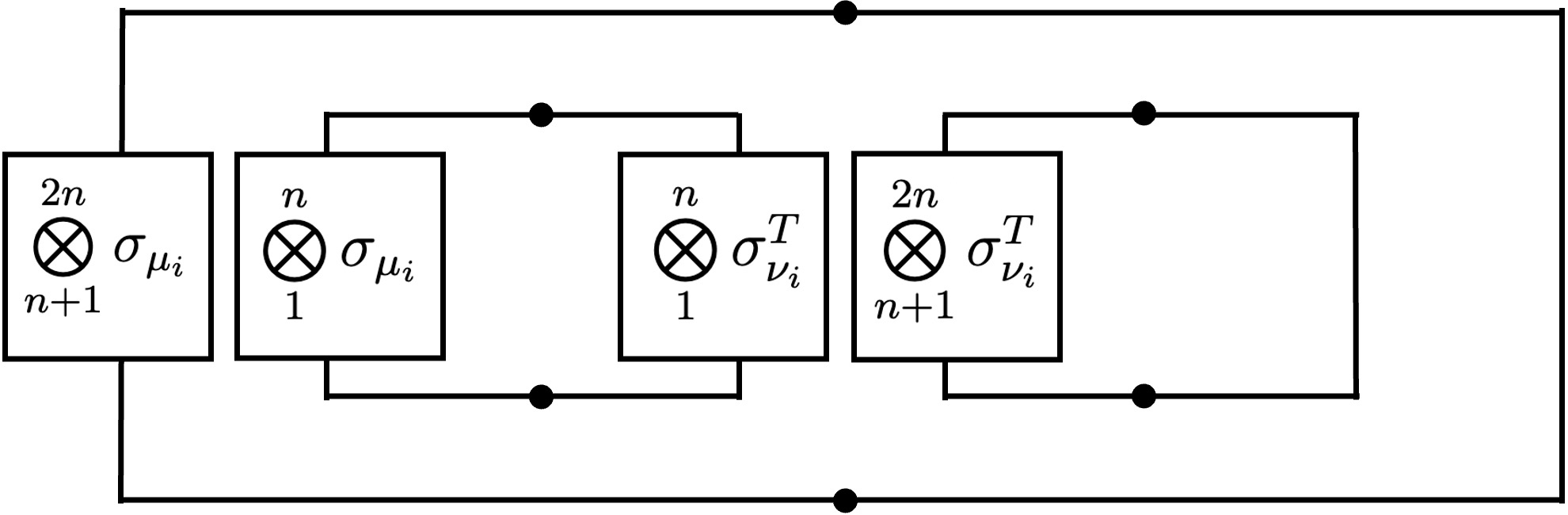}
    \\
    & = \frac{1}{2^{3 n}}\Tr(\otimes_{i=n+1}^{2n} \sigma_{\mu_i})\Tr(\otimes_{i=n+1}^{2n} \sigma_{\nu_i})\Tr(\otimes_{i=1}^{n} \sigma_{\mu_i}\sigma_{\nu_i}) 
    \\
    & = \Pi_{i=1}^{n} \frac{1}{2}\Tr(\sigma_{\mu_i}\sigma_{\nu_i})\Pi_{i=n+1}^{2n} \frac{1}{2^2}\Tr(\sigma_{\mu_i})\Tr(\sigma_{\nu_i})
    \\
    & = \delta_{\bar \mu \bar \nu} \Theta_{AB}(\sigma_{\bar \nu}) \Theta_{AB}(\sigma_{\bar \mu}),
\end{split}
\end{equation}
where $\Theta_{AB}(\sigma_{\bar \mu})$ equals $1$ if $\sigma_{\bar \mu}$ has support \emph{only} on the EPR pairs between $A$ and $B$, \ie on the first $n$ sites, and $0$ otherwise. Therefore, 
\begin{align}
     & C_{i} = -\sum_{\bar \mu \in A_B} c_{i,\bar \mu}^2 e^{-i g \size(\sigma_{\bar \mu})/{n}},
\end{align}
where the sum runs over the operators in the expansion \eqref{opExp} that have support only on the qubits of $A$ that are entangled with $B$, and $F_{\text{pure}}$ is given by 
\begin{align} \label{FLowerB}
     F_{\text{pure}} = \frac{1}{16}\Bigl\vert1 -\sum_{i=1}^3 \sum_{\bar \mu  \in A_B} c_{i,\bar \mu }^2 \, e^{-i g \size(\sigma_{\bar \mu })/n }\Bigr\vert^2.
     \end{align}
Let us consider for the moment the two-boundary case where no qubits of $A$ or $B$ are entangled with $C$, \ie when $A_B=A$. At large times, when the operator has scrambled throughout the entire system $A$, the teleportation succeeds due to the peaked-size mechanism \cite{Schuster:2021uvg}, mentioned in the introduction. In this case, the probability distribution of \eqref{eq:c_iNormalization} is effectively uniform over all possible Pauli strings. For a uniformly random Pauli string, each site independently carries the identity with probability $1/4$ and a non-identity Pauli operator with probability $3/4$. Therefore, the size distribution is effectively binomial over $n$ trials with success probability $3/4$. This gives that the mean of $\size(\sigma_{\bar \mu})/n$ is $3/4$ where the standard deviation goes like $1/\sqrt{n}$.
Hence, in the two-boundary case, with large $n$ and for large times, the phases in the sum of \eqref{FLowerB} can be replaced with a global phase that is given by the mean, which together with Eq.~\eqref{eq:c_iNormalization}, results in 
\begin{equation} \label{FpureEst}
    F_{\text{pure}} \approx \Bigl\vert\frac{1}{4}(1 - 3 e^{-i g \bar S/n})\Bigr\vert^2 \approx \Bigl\vert\frac{1}{4}(1 - 3 e^{-i g\frac{3}{4}})\Bigr\vert^2.
\end{equation}
Choosing $g =  4\pi/3$ therefore yields perfect teleportation.
This late-time perfect fidelity is ruined when system $C$ is introduced. In this case, $ \sum_{\bar \mu  \in A_B} c_{i,\bar \mu }^2$ is the probability of $\sigma^{A_r}_{i,t}$ not having support on the $A_C$ sites, and this probability decreases exponentially with the size of $A_C$ at late times.  Let us, then, consider the fidelity before the operator has scrambled over the entire system.

\begin{figure}
    \centering
    \includegraphics[width=0.45\textwidth]{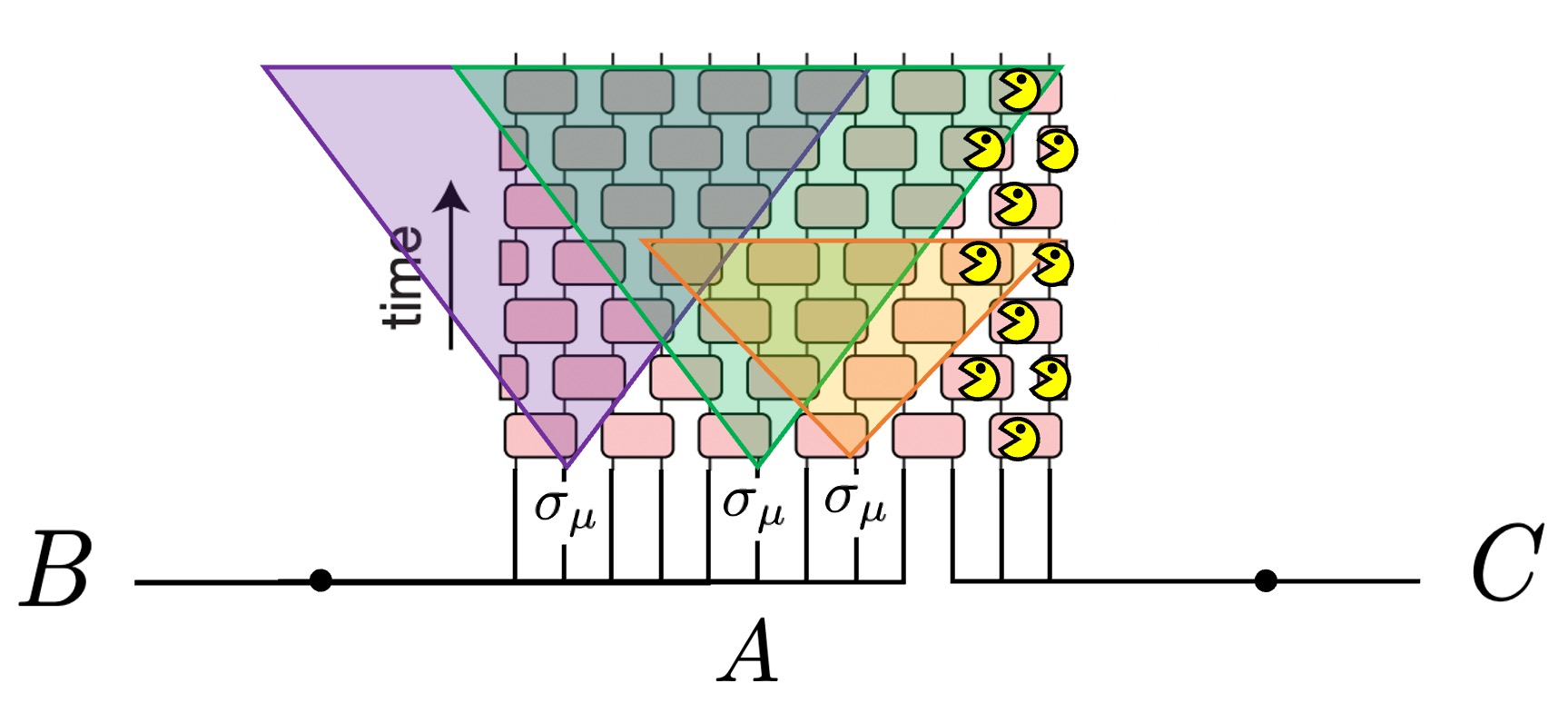}
    \caption{The operators spread with time, allowing for the peaked size mechanism to work. The regions to which the operators can spread without damaging the fidelity are limited in $1D$ by the distance to the sites that are entangled with $C$.}
    \label{fig:yummy}
\end{figure}

If the unitary evolution is strictly causal, then $\sigma^{A_r}_{i,t}$ has support only within the light cone, \ie on sites $A_j$, $|j-r|\leq t$.\footnote{Here we use units in which the causal velocity is one. In the circuit setting, $t$ corresponds to the number of circuit layers.} 
More generally, the Lieb-Robinson bound ensures that this is true, up to exponentially small corrections, also for systems without strictly causal dynamics  \cite{lieb1972finite, xu2024scrambling}.
In many chaotic local systems, there is an effective butterfly velocity $v_b$ that sets a light cone where the operator has scrambled. Recall that the sites that are entangled with $C$ are in $A_{j>{n}}$, and let us define their minimal distance from the site $r$ in which the information was injected, as
\begin{equation}\label{eq:Delta}
    \Delta=\min(r,n-r).
\end{equation} 
Therefore, after a critical time 
\begin{equation} \label{eq:tc}
  t_c\equiv \frac{\Delta} {v_b}  ,
\end{equation} 
which is when the initial information has scrambled to sites in $A$ that are entangled with $C$, we expect to see a very fast drop in the fidelity (see Fig. \ref{fig:yummy}).

For times $\Delta<v_bt<n-\Delta$,
only one front of the light cone has
reached the $A_C$ region. Heuristically, if the time evolution of the operator injected in $A$ is a random superposition of different Pauli strings $\sigma_{\bar \mu}$ contained within the light cone, then, after time $t_c$, the probability of $\sigma_{\bar \mu}$ to have support only in $A_{j<{n}}$, is around 
$4^{\Delta -v_b t}$.  
This is because there are $4$ Pauli operators that can be inserted in each  
of the $\Delta-v_bt$ sites that are inside the light cone but entangled with $C$, and only one option is the identity operator.
Therefore, roughly only  
$4^{\Delta -v_b t}$ of the terms in \eqref{FLowerB} will contribute, and those will have an average size of  $3/4(v_bt+\Delta)$.\footnote{This is because one side of the light cone continues to grow with time and contributes $3v_b t/4$ to the size while the other side has reached the $A_C$ sites, and therefore only contributes $3\Delta/4 $.} Thus, after  $t_c$,\footnote{At $v_bt\simeq n-\Delta$, the operator light cone also reaches
the $A_C$ region through the second interface of $A_B$ and $A_C$, as there are periodic boundary conditions. By this time, the
contribution retained in the estimate above has already been
suppressed by a factor $4^{-(n-2\Delta)}$. Thus, when
$n-2\Delta$ is larger than a few sites, including the second front
only modifies an already exponentially small contribution.} 
\begin{equation}
    F_{\text{pure}} \approx  \Bigl\vert\frac{1}{4}\Bigl(1 - 3 \frac{e^{-i  \frac{3 g}{4n}(v_bt+\Delta)}}{4^{v_b t-\Delta}}\Bigr)\Bigr\vert^2,\, \qquad \text{for  }~ t>t_c.
\end{equation}
If we choose the value of $g$ that optimizes the above expression for $F_{\text{pure}}$, we obtain 
\begin{equation} \label{Fbound}
    F_{\text{pure}} \approx  \Bigl\vert\frac{1}{4}\Bigl(1 +  \frac{3}{4^{v_b t-\Delta}} \Bigr)\Bigr\vert^2.
\end{equation}
We see that the fidelity drops exponentially fast after the light cone reaches $A_C$.

The way around this is to look at intermediate times, such that the operators have not had the chance to spread to the $A_C$ entangled qubits, \ie times smaller than $t_c$, cf. Eq.~\eqref{eq:tc}. In this case, the average size of the operators, $S(\sigma_{\bar \mu})/n$ will be roughly $\frac{2 v_b t}{n} \frac{3}{4}$ and, 
\begin{equation} \label{FpureApprox}
    F_{\text{pure}} \approx  \Bigl\vert\frac{1}{4}(1 - 3 e^{-i g \frac{2 v_b t}{n} \frac{3}{4}} )\Bigr\vert^2.
\end{equation}
Choosing $g = \pi \frac{2 n}{3 v_b t }$ again gives perfect fidelity. 
Notice, however, that earlier times require a larger $g$.\footnote{One might be worried that $g$ scales with $n$, and would become large for large systems. However, recall from Eq.~\eqref{eq:V} and \eqref{interactionHam} that the coupling involving the different $A-B$ EPR pairs is also rescaled as $1/n$ and therefore the coupling per pair scales like $1/(v_b t)$. Furthermore, at times of the order of $t_c$ the coupling per pair is of order $1/\Delta$, so if  $\Delta$ scales with $n$ the coupling per pair is even smaller (of order $1/n$).} 
As the fidelity starts to drop at time $t_c$, because of the effect of the $A-C$ entangled pairs, the minimal value of $g$ needed for good fidelity is around
\begin{equation}\label{gmin}
    g_{\min} =\pi \frac{2n }{3\Delta },
\end{equation}
which grows as the information is injected in sites that are getting closer to the $A_C$ entangled qubit sites.

In models with a gravitational dual, the coupling strength $g$ is associated with the amount of negative energy generated in the bulk. This negative-energy perturbation in turn controls the magnitude of the null shift of excitations propagating through the wormhole, allowing them to reach the receiving boundary \cite{gao2017traversable,maldacena2017diving,freivogel2020traversable,brown2023quantum}. In a multi-boundary wormhole, such as those depicted in Figs.~\ref{fig:causal shadow}-\ref{fig:PenroseCS}, the minimum null shift required for successful teleportation depends on the locations of the insertion and recovery points. If the corresponding bulk trajectory penetrates more deeply into the causal shadow, for example by passing closer to the region associated with the third boundary, it must undergo a larger null shift to emerge at the target boundary. Consequently, a larger-magnitude negative-energy perturbation, and hence a stronger coupling $g$, is required. This mirrors the behavior of our protocol, in which the minimum coupling increases as the injection point approaches the region entangled with the third party. We stress that this is just an analogy. The model at hand is just assumed to be scrambling, but is not holographic.

A potential caveat regarding the holographic analogy is that in the infinite temperature limit, the distance between the horizons across the causal shadow region is expected to shrink to zero because of the bipartite nature of the entanglement in $\ket{\Psi}$ \cite{peach2017tensor}.
In holography, a finite causal shadow is instead associated with multipartite, rather than purely bipartite entanglement. 
Although multipartite entanglement is absent from our initial state, it can emerge dynamically through scrambling. In particular, at times $v_b t \geq \Delta$, when the teleportation fidelity begins to decrease, the evolved message operator $\sigma^{A_r}_{\nu,t}$
acquires support on both $A_B$ and $A_C$. For generic scrambling dynamics, the evolved state $\sigma^{A_r}_{\nu,t}\ket{\Psi}$ is then expected to develop multipartite entanglement involving degrees of freedom associated with all three parties.

\section{Random circuits}\label{sec:radncir}
In this section, we specialize to random-circuit dynamics. Random circuits provide analytically
and numerically tractable models of chaotic many-body evolution. After
averaging over the gates, they reproduce universal features of
scrambling, such as operator growth and the propagation and broadening
of operator fronts, while suppressing model-specific microscopic
details \cite{nahum2018operator}. We consider circuits composed of two-qubit gates drawn
independently from the Haar measure, arranged either in a $1D$ 
brickwork geometry or in an all-to-all architecture, which we refer to as $0D$. The time parameter $t$ will then be interpreted as the number of circuit layers. The brickwork
circuit allows us to resolve where the information spreads relative to
the sites entangled with the third party, while the all-to-all circuit
provides a model of fast scrambling.

However, studying random or chaotic circuits is infeasible for large systems. 
Fortunately, there is a computationally tractable workaround. The Clifford group on $n$ qudits of local dimension $q$ is a unitary $2$-design when $q$ is a prime number \cite{nahum2018operator, dankert2009exact}. This means that the Haar average of circuit quantities which have no more than two copies of $U$ and two copies of $U^\dagger$ can be obtained by averaging instead over the Clifford group with uniform distribution. For systems of qubits, this replacement works even for quantities, like $F_{\text{EPR}}$ in Eq.~\eqref{eq:FEPR3U}, that have three copies of $U$ (or $U^T$) and three copies of $U^\dagger$ (or $\bar U$) \cite{webb2015clifford, kueng2015qubit, zhu2017multiqubit}.

Clifford circuits are special in that they take a string of Pauli operators to a single different string, without generating superposition, and they are easy to simulate classically. 
Therefore, when considering Clifford circuits, the evolved Pauli operators defined in \eqref{evolvedPauli} can be expressed as
\begin{equation}\label{sigArRand}
    \sigma^{A_r}_{\mu,t} =\bigotimes_{k=1}^{2n}\sigma_{\mu_{k,t}}\otimes \boldsymbol{1}_B \otimes \boldsymbol{1}_C,
\end{equation}
where we ignore a possible overall minus sign as it won't change the final results. 
Similarly, using Eq.~\eqref{tilsigBr}, we obtain
\begin{equation}\label{sigBrRand}
\begin{split}
\tilde{\sigma}_{\mu,t}^{B_r} & =(-1)^{1+\delta_{\mu,0}}(\sigma_{\mu,t}^{B_r})^T 
\\
& = (-1)^{1+\delta_{\mu,0}}\boldsymbol{1}_A \otimes \bigotimes_{k=1}^{2n}\sigma_{\mu_{k,t}}^T\otimes \boldsymbol{1}_C.
\end{split}
\end{equation}

Substituting the above relations into Eq.~\eqref{fidelityExact} and using Eq.~\eqref{VsAction}, we find that
\begin{equation}
\begin{split}
\label{fidelityExact2}
    \fepr  &=\frac{1}{2^4}\sum_{\nu,\mu}(-1)^{\delta_{\nu,0} +\delta_{\mu,0}}e^{\frac{ig}{n}(\size(\sigma^{A_r}_{\nu,t})- \size(\sigma^{A_r}_{\mu,t}))}\\
    &\times
    \bra{\Psi} \sigma^{A_r}_{\nu,t} (\sigma^{B_r}_{\nu,t})^T  (\sigma^{B_r}_{\mu,t})^T \sigma^{A_r}_{\mu,t}\ket{\Psi}.
\end{split}
\end{equation} 
Notice that when $\sigma^{B_r}_{\mu,t}$ and $\sigma^{A_r}_{\mu,t}$ are Pauli strings of qubit Pauli matrices, so are the products $(\sigma^{B_r}_{\nu,t})^T  (\sigma^{B_r}_{\mu,t})^T$ and $\sigma^{A_r}_{\nu,t} \sigma^{A_r}_{\mu,t}$ (up to a phase),  
and therefore $\bra{\Psi} \sigma^{A_r}_{\nu,t} (\sigma^{B_r}_{\nu,t})^T  (\sigma^{B_r}_{\mu,t})^T \sigma^{A_r}_{\mu,t}\ket{\Psi}$ has the form given in \eqref{expectToTrace}. 
Thus,
\begin{equation}\label{expectFour}
\bra{\Psi} \sigma^{A_r}_{\nu,t} (\sigma^{B_r}_{\nu,t})^T  (\sigma^{B_r}_{\mu,t})^T \sigma^{A_r}_{\mu,t}\ket{\Psi}=T^{AB}_{\mu\nu}T^{BC}_{\mu\nu}T^{CA}_{\mu\nu},
\end{equation} with 
\begin{align}
&T^{AB}_{\mu\nu}\coloneqq\prod_{k=1}^{n}\frac12\Tr[\sigma_{\mu_{k,t}}\sigma_{\nu_{k,t}}\sigma_{\nu_{k,t}}\sigma_{\mu_{k,t}}] =1,\\
&T^{BC}_{\mu\nu}\coloneqq\prod_{k=n+1}^{2n}\frac12\Tr[\sigma_{\mu_{k,t}}\sigma_{\nu_{k,t}}] =\prod_{k=n+1}^{2n}\delta_{\mu_{k,t},\nu_{k,t}},\\
&T^{CA}_{\mu\nu}\coloneqq\prod_{k=n+1}^{2n}\frac12\Tr[\sigma_{\nu_{k,t}}\sigma_{\mu_{k,t}}] = T^{BC}_{\nu\mu},
\end{align}
where we have used that $\sigma_{\mu_k} \sigma_{\mu_k}=\boldsymbol{1}_2$ and the trace is cyclic.\footnote{When using \eqref{expectToTrace} for the above derivation, each $\sigma_{\mu_i}$ is the product of two Pauli matrices, either $(\sigma^{B_r}_{\nu,t})^T  (\sigma^{B_r}_{\mu,t})^T$ or $\sigma^{A_r}_{\nu,t} \sigma^{A_r}_{\mu,t}$, originating from \eqref{expectFour}. Hence, the double number of Pauli matrices in the traces.}
$T^{AB}_{\mu\nu}$ arises from the parts of the $\sigma^{A_r}_{\mu,t}$ and $\sigma^{B_r}_{\mu,t}$ Pauli strings that are on sites that are EPR paired between $A$ and $B$. On the other hand, $T^{BC}_{\mu\nu}$ and $T^{CA}_{\mu\nu}$ are from the parts of the Pauli strings that are on sites that are EPR paired between $B$ and $C$, and $A$ and $C$, respectively. 
Since $\Tr[\sigma_\mu \sigma_\nu] = 2\delta_{\mu\nu}$, the last two expressions imply that $\fepr$ gets non-zero contributions only when $\sigma^{A_r}_{\nu,t}  \sigma^{A_r}_{\mu,t}$ acts trivially on $A_C$, the qubits that are entangled with $C$ (equivalently, when $\sigma^{B_r}_{\nu,t}  \sigma^{B_r}_{\mu,t}$ acts trivially on $B_C$). 
Finally, we arrive at
\begin{equation}\label{FClifFinal}
\begin{split}
    \fepr &=\frac{1}{2^4}\sum_{\mu,\nu}(-1)^{\delta_{\mu,0} +\delta_{\nu,0}} \Theta_{AB}(\sigma^{A_r}_{\nu,t} \sigma^{A_r}_{\mu,t})\\
    &\times e^{ \frac{i g}{n} \left( \size(\sigma^{A_r}_{\nu,t})- \size(\sigma^{A_r}_{\mu,t})\right)}.
\end{split}
\end{equation}
Taking the average of the above over randomly chosen Clifford gates will give the result for the random circuit.

Notice that the fidelity is upper-bounded by
\begin{equation} \label{upperbound}
\begin{split}
    \fepr & \leq \frac{1}{2^4}\sum_{\mu,\nu} \Theta_{AB}(\sigma^{A_r}_{\nu,t} \sigma^{A_r}_{\mu,t})
    \\
    & = \frac{1}{4}+\frac{1}{4}\sum_{i=1}^3\Theta_{AB}(\sigma^{A_r}_{i,t}),
\end{split}
\end{equation}
where the equality on the second line was derived using the fact that $\sigma^{A_r}_{\nu,t}\sigma^{A_r}_{\mu,t}=1$ for $\mu=\nu$, and $\sigma^{A_r}_{i,t}\sigma^{A_r}_{j,t}=i\varepsilon_{ijk}\sigma^{A_r}_{k,t}$ for $i\neq j$, where the imaginary phase does not affect $\Theta_{AB}$. Therefore, the fidelity drops as soon as the operators spread outside of the $AB$ EPR pairs.

\subsection{1D}\label{sec:radncir1d}

We now focus on a random brickwork layered circuit with periodic boundary conditions for each individual system.
We start by considering the probability of $\sigma^{A_r}_{j,t}$ having support at each site of the chain as a function of time. In what follows, time represents the number of circuit layers. In \cite{nahum2018operator}, it was shown that the OTOC with $\sigma^{A_r}_{j,t}$ and $\sigma^{A_x}_{j,0}$, which is related to the probability that $\sigma^{A_r}_{j,t}$ has support at site $x$,\footnote{The OTOC is proportional to $\Tr([\sigma^{A_r}_{j,t},\sigma^{A_x}_{j,0}]^2)$. Let us expand $\sigma^{A_r}_{j,t}$ as a sum over Pauli strings, as in \eqref{opExp}. Notice that only terms in the expansion that have at site $x$ Pauli matrices other than $1$ and $\sigma_j$ contribute to $[\sigma^{A_r}_{j,t},\sigma^{A_x}_{j,0}]$, and $\Tr([\sigma^{A_r}_{j,t},\sigma^{A_x}_{j,0}]^2)$ is proportional to the sum of $c_{i,\bar \mu}^2$ over all such terms. This is exactly the probability of $\sigma^{A_r}_{j,t}$ having a Pauli matrix other than $1$ and $\sigma_j$ at site $x$, which for a scrambling evolution, is $2/3$ the probability of it having support at site $x$.} should saturate for sites that satisfy $|x-r|<\frac{3}{5}t$. For sites outside this range, the OTOC drops to zero as a function of the site location, with width $\Delta x = \frac{4}{5}\sqrt{t}$. 
As the support probability of the expanding message constrains the fidelity of our protocol, cf. Eq.~\eqref{upperbound}, we expect the above scales to dictate the behavior of the protocol as the message spreads to the sites entangled with the third party. More precisely, if $\Delta$ is the distance between the location of the first site that is entangled with $C$, and the site where we inject the message, the above suggests that after a time $3 t+4 \sqrt{t}=5\Delta$, the fidelity will start to decrease.
At around $t = 5/3\Delta$, the decrease will become exponential, and the fidelity will converge quickly to $1/4$, as can be inferred from Eqs.~\eqref{FClifFinal}-\eqref{upperbound}.

To test this behavior, we perform a numerical simulation of the averaged protocol with the brickwork Clifford circuits described in the previous section.
However, to reduce the simulation time, we turn away from the symmetric setup, and instead, take systems $A$ and $B$ to each have $1000$ qubits while system $C$ will contain $40$ qubits.
In the initial state, each of the first $980$ qubits in $A$ is EPR paired with the corresponding qubit in $B$, while each qubit of $C$ is entangled with one of the last $20$ qubits of $A$ or $B$. The parameter $n$, which previously denoted half the qubits in each system, will now be equal to $980$, and will represent the qubits that are entangled between $A$ and $B$.

Although different from the symmetric analysis presented so far, the generalization is straightforward by setting systems $A$ and $B$ to have $n+n_c$ qubits instead of $2n$, and keeping the coupling to be between the first $n$ qubits of $A$ and $B$ as before. In this case, we expect the results to have no significant dependence on the size $2n_c$ of system $C$, as long as system $C$ contains at least a few qubits. This is because the fidelity drops exponentially fast when the operator $\sigma^{A_r}_{i,t}$ has spread to the qubits that are entangled between $A$ and $C$, as evident from Eq. \eqref{Fbound}.\footnote{To be more precise, when the light cone reaches $A_C$ at $t_c=5/3\Delta$, there is already a region in $A_C$ of size $\Delta x = 4/5\sqrt{t_c} = 4\sqrt{\Delta}/\sqrt{15}$ to which the operator has spread with nonvanishing probability. If this region is smaller than $n_c$, then increasing $n_c$ will have an exponentially small effect on the protocol.}

\begin{figure*}
    \centering
    \subfloat[]{\includegraphics[width=0.45\textwidth]{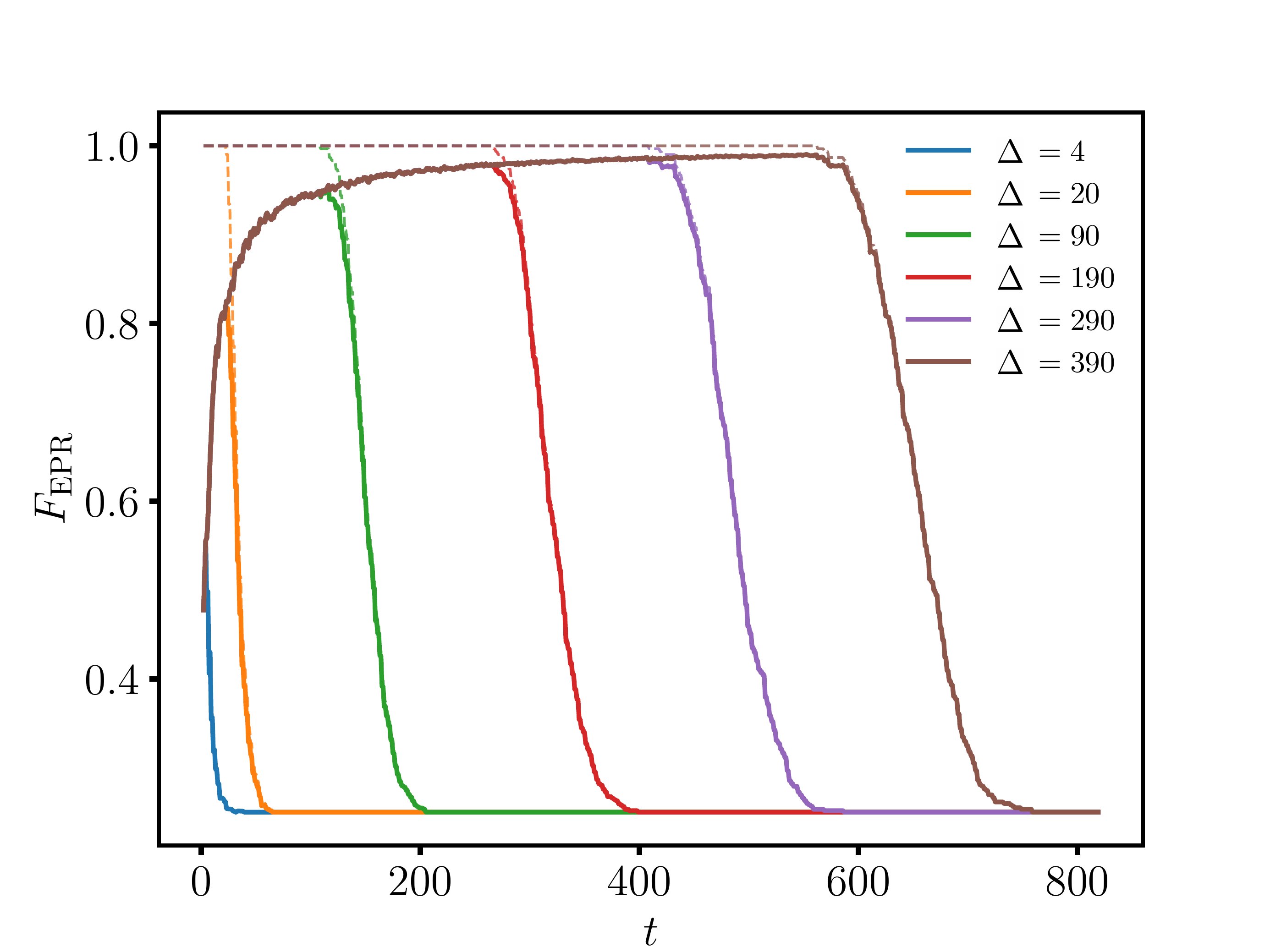}\label{fig:subfig_a}}
    \subfloat[]{\includegraphics[width=0.45\textwidth]{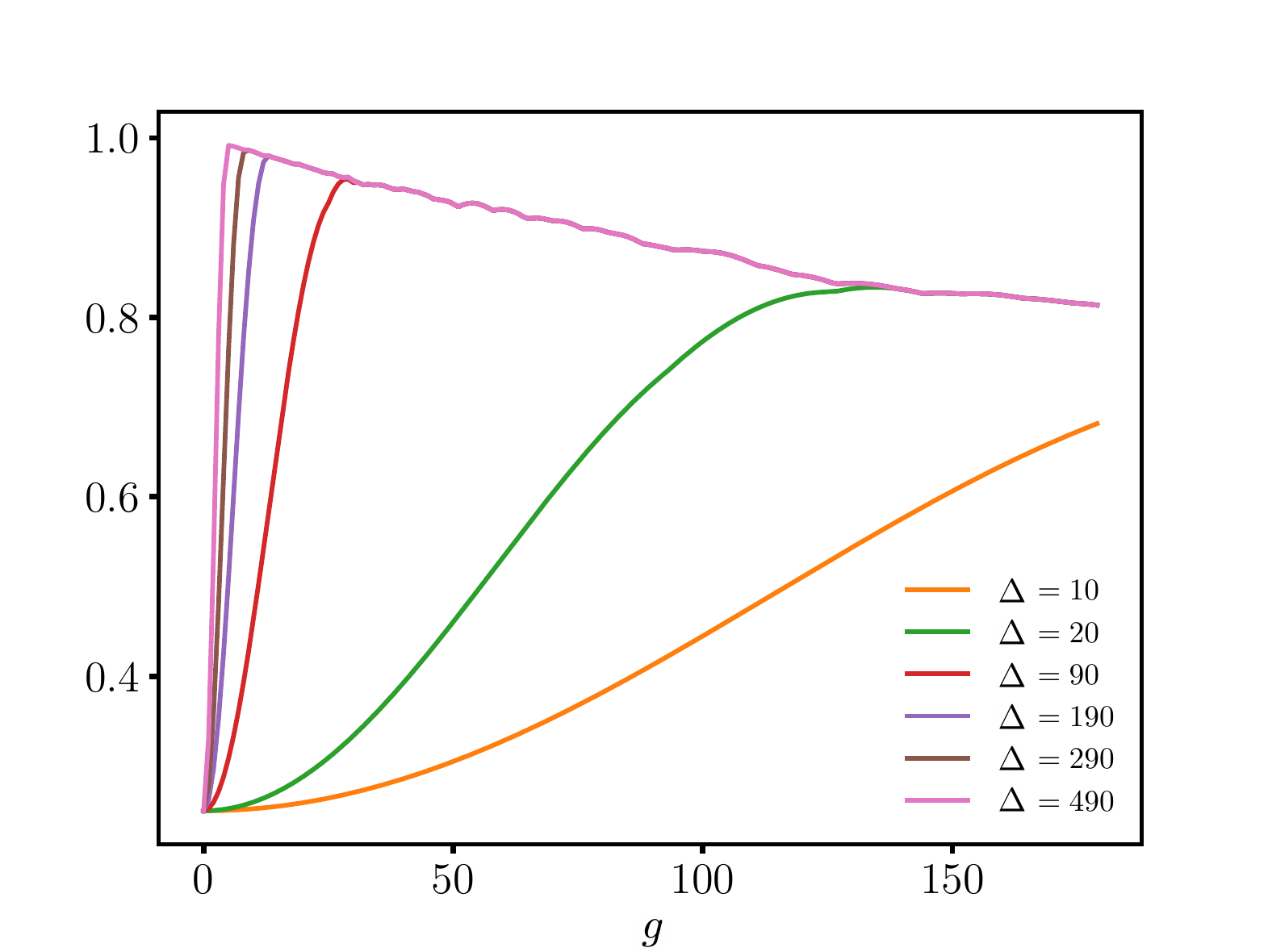}\label{fig:subfig_b}}
    \caption{
    The teleportation fidelity for several distances $\Delta$
between the insertion site $r$ and the region of $A$ that is entangled
with system $C$. System $A$ contains $1000$ qubits: the first
$n=980$ are EPR paired with $B$, and the remaining $20$ are EPR paired
with $C$. The dynamics are generated by a one-dimensional
brickwork random Clifford circuit with periodic boundary conditions,
and the results are averaged over $152$ circuit realizations.
\textbf{(a)} Fidelity as a function of the number of circuit layers
$t$, with
$g(t)=2\pi n/(3v_b t)$ (see discussion under Eq.~\eqref{FpureApprox}) and $v_b=3/5$. The dashed curves correspond to the upper
bound in Eq.~\eqref{upperbound}. The initial growth results from the
decreasing variance of the coupling-induced phase, while the decay
begins when the broadened operator front reaches the qubits entangled
with $C$. The onset of decay and rapid-decay timescales approximately agree with the values described in the main text, namely $3t+4\sqrt{t}\simeq5\Delta$ and
$t\simeq5\Delta/3$, respectively. 
\textbf{(b)} The maximal fidelity over the protocol time $t$, plotted
as a function of the coupling $g$. The maxima occur near
$g_{\min}=2\pi n/(3\Delta)$ as predicted by Eq.~\eqref{gmin}. Their slight displacement toward larger
$g$ results from the
broadening of the operator front. The decrease in the fidelity with $g>g_{\min}$ is due to the increasing variance of the coupling-induced phase, which is an outcome of the earlier optimal time involved.}
\end{figure*}

In Fig.~\ref{fig:subfig_a}, we present the fidelity obtained from
Eq.~\eqref{FClifFinal} as a function of time for several distances
$\Delta$ between the insertion site and the region $A_C$ that is
entangled with system $C$. The fidelity is averaged over $152$ random realizations of the circuit. The solid curves show the fidelity, while
the dashed curves show the upper bound in Eq.~\eqref{upperbound}. At
early times, before the operator front reaches $A_C$, the curves for
the different distances coincide. During this regime, the operator
scrambles over a region of length approximately $2v_b t$, and its mean
size is
$
\overline{\size(\sigma^{A_r}_{i,t})}
\simeq \frac{3}{4}(2v_b t).
$
We therefore choose the coupling according to the mean-size estimate,
$
g(t)=\frac{2\pi n}{3v_b t},
$ so that the mean coupling-induced phase is equal to $\pi$, as in the
discussion around Eq.~\eqref{FpureApprox}.

The initial increase in fidelity reflects the progressive narrowing of
this phase distribution. In particular,
$\operatorname{Var}[g\size(\sigma^{A_r}_{i,t})/n]\sim 1/t$ for
$t\ll t_c$.\footnote{Before the operator reaches $A_C$,
we may approximate it as a random Pauli string on
$2v_b t$ sites inside the light cone. With this approximation, as in the discussion below Eq. \eqref{FLowerB}, the size is drawn from a Binomial distribution, with $2v_b t$ trials and $3/4$ probability of trial success, having a variance that scales as $t$. Therefore, with $g\sim n/t$, the variance of $gS_{AB}/n$ scales as $1/t$.}
Therefore, the peaked-size approximation, which allows $\size(\sigma^{A_r}_{i,t})$ to be treated as a constant that does not depend on $i$, becomes more accurate with
time.

This improvement competes with the propagation of the operator into
$A_C$. As discussed at the beginning of the subsection, the leading edge of the broadened front reaches $A_C$ when
$3t+4\sqrt{t}\simeq 5\Delta$, at which point the upper bound begins to
decrease. The light cone reaches $A_C$ around
$t_c\simeq \Delta/v_b=5\Delta/3$, producing the rapid decay toward
$F_{\mathrm{EPR}}=1/4$. These two times can be seen in Fig.~\ref{fig:subfig_a}, where one time sets the beginning of the decay of each curve, and the other sets the decay width. Increasing $\Delta$ translates the decay to
later times without changing the common initial growth. Thus, at fixed
absolute distance $\Delta$, this behavior is insensitive to the total
system size, provided that $n$ and $n_c$ are sufficiently
large.

As previously mentioned, the three-boundary wormhole is expected to
have a causal shadow, so that the distance that must be traversed
depends on the angular position from which the information is sent.
An analogous dependence is visible in
Fig.~\ref{fig:subfig_b}, which shows, for each value of $g$, the
fidelity \emph{maximized over the protocol time $t$}, for several distances
$\Delta$.

Before the operator reaches $A_C$, the mean coupling-induced phase is
approximately
$
\frac{g\,\bar S_{AB}}{n}
\simeq
\frac{3g v_b t}{2n}.
$
For fixed $g$, the first optimal time is therefore
$
t_{\mathrm{opt}}(g)
\simeq
\frac{2\pi n}{3v_b g},
$
which is inversely proportional to $g$. However, this time is
constrained by $t_c\simeq\Delta/v_b$, since the fidelity begins to
decrease once the operator reaches $A_C$. For $g<g_{\min} = \frac{2\pi n}{3\Delta}$, the
coupling-induced phase cannot reach $\pi$ before this happens.
Increasing $g$ therefore increases the maximal fidelity until
$ g_{\min},$
as given in Eq.~\eqref{gmin}. The peaks in
Fig.~\ref{fig:subfig_b} occur slightly above this estimate because
Eq.~\eqref{gmin} treats the operator front as sharp, whereas the
leading tail of the broadened front reaches $A_C$ earlier. Choosing a
somewhat larger $g$, and hence a somewhat earlier
$t_{\mathrm{opt}}$, reduces this leakage into $A_C$.

If the operator size were perfectly sharp, the approximation in
Eq.~\eqref{FpureApprox} would predict unit fidelity for any
$g\gtrsim g_{\min}$, since one could simply choose
$t=t_{\mathrm{opt}}(g)$. In contrast, the numerical curves decrease
for $g$ above the peak. This is a finite-size effect associated with
the width of the operator-size distribution. A larger coupling
requires an earlier optimal time, at which the phase distribution is
broader, and the peaked-size mechanism is less accurate. Expanding
Eq.~\eqref{FClifFinal} around the optimal mean phase shows that the infidelity
$
1-F_{\mathrm{EPR}}
\propto
\operatorname{Var}\!\left[
\frac{g\size(\sigma^{A_r}_{i,t_{\mathrm{opt}}})}{n}
\right]
\propto
\frac{g}{n}.
$
At the peak, where $g\simeq g_{\min}\sim n/\Delta $, this gives
$
1-F_{\mathrm{EPR}}^{\max}
=
O\!\left(\frac{1}{\Delta}\right).
$
The numerical peak heights in  Fig.~\ref{fig:subfig_b} are consistent with this scaling. 

In the large $n$ limit, one may be interested in the maximal fidelity as a function of the ratio $\Delta/n$, instead of $\Delta$. For a fixed ratio and $g>g_{\min}$, the infidelity for this ratio vanishes as
$n\rightarrow\infty$. 
Thus, in the thermodynamic limit, the decay
above the peak is replaced by an approximately unit-fidelity plateau,
and the maximal fidelity approaches unity whenever the distance from
the $C$-entangled region grows with the system size.

\subsection{0D (fast scrambling)}\label{sec:radncir0d}

We now replace the brickwork circuit by a $0D$ 2-local circuit. At
each time step, the sites are divided into randomly chosen pairs, and
a random two-qubit unitary acts on each pair. We retain the notation
introduced in the previous subsection: systems $A$ and $B$ each
contain $n+n_c$ qubits, of which $n$ are EPR paired between $A$ and
$B$, while the remaining $n_c$ qubits in each system are EPR paired
with $C$. We denote by $A_C$ the set of $n_c$ qubits in $A$ that are EPR paired with system $C$.
The two-boundary case, obtained by setting $n_c=0$, was analyzed in
Refs.~\cite{Schuster:2021uvg,nezami2023quantum}.

\begin{figure*}
    \centering
    \subfloat[]{\includegraphics[width=0.45\textwidth]{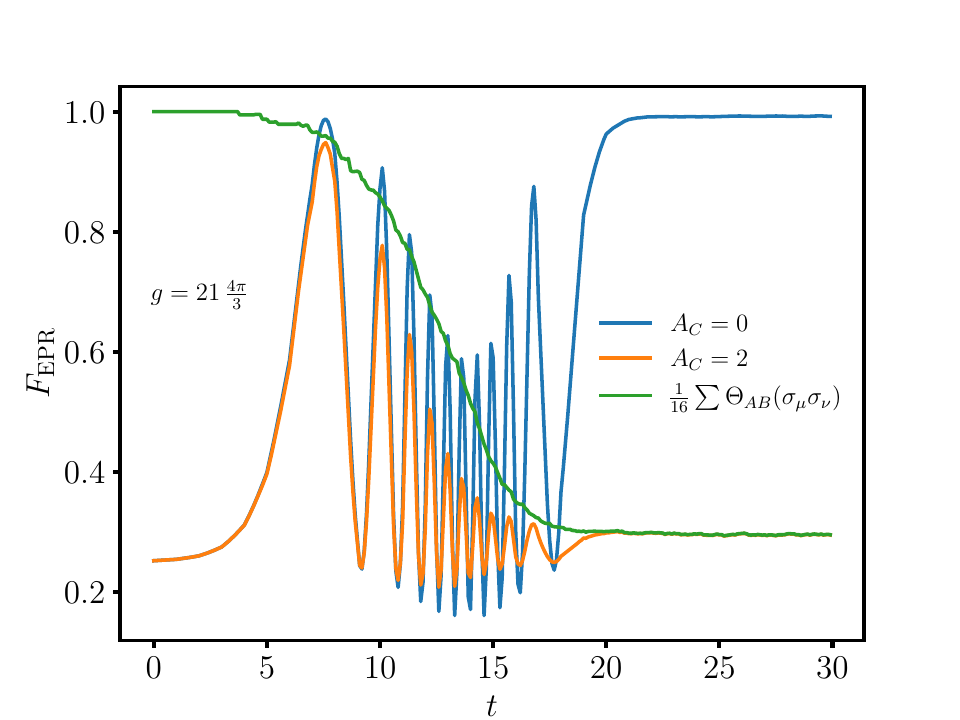}\label{fig2:subfig_a}}
    \subfloat[]{\includegraphics[width=0.45\textwidth]{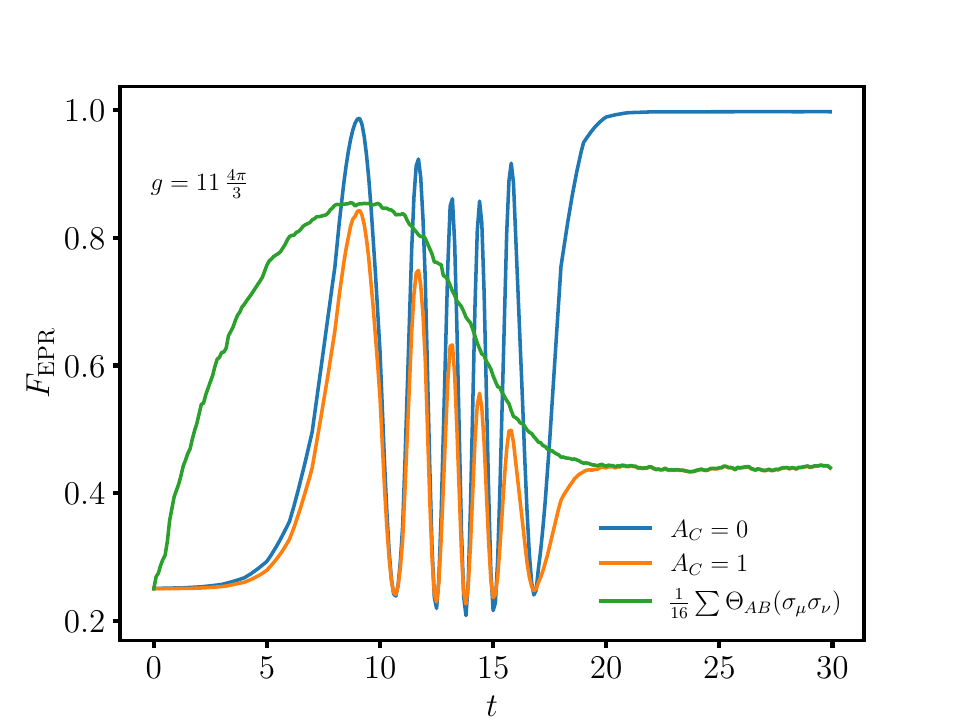}\label{fig2:subfig_b}}
    \caption{The fidelity as a function of time for the $0D$ dynamics. System $A$ has $n +n_c=10^5$ sites, of which $n_c$ sites are EPR paired with system $C$ and the remaining $n$ are EPR paired with system $B$. The data is averaged over $4$ 
    different random pairings of the sites in the evolution. For each choice of pairing, we average over $\sim 70$ random selections of the Clifford gates.
    The data is then averaged again over 15 different locations for the $A_C$ sites. 
    The encoding unitary maps the three Pauli matrices to $p=100$ sites. The green curve is the upper bound of  Eq.~\eqref{upperbound}. Its value is controlled just by the averaged support within the $A_C$ sites. \textbf{(a)} $n_c=2$ sites in $A_C$, located outside of the $p$ sites. $g=21 \frac{4 \pi}{3}$ \textbf{(b)} A single site in $A_C$ ($n_c=1$), located within the $p$ sites. $g=11 \frac{4 \pi}{3}$. The values of $g$ are selected to have the peak of the orange curve relatively close to the maximum of the green upper bound. The late-time limit matches Eq.~\eqref{upperbound}, with $\Theta_{AB}(\sigma_{i,t}^{A_r})$ replaced by $4^{-n_c}$, the expected value after the $\sigma_{i,t}^{A_r}$ has fully scrambled within the system.}
    \label{F0doft}
\end{figure*}

As explained in \cite{Schuster:2021uvg}, both the averaged operator size and its standard deviation grow exponentially as $e^{\lambda t}$, where $\lambda$ is the Lyapunov exponent and is equal to $\log(\frac{8}{5})$ for qubits.
Having a separation between the average and standard deviation is crucial for the peaked size mechanism, as explained around \eqref{FpureEst}. To rectify this, the authors in \cite{Schuster:2021uvg} encode the initial information over $p$ sites, such that there is an encoding unitary $E$ which maps each of the three Pauli matrices that at $t=0$ act on site $r$, to a Pauli string with support on $p$ sites. If $p$ is much smaller than the number of sites in each system, then the size of each Pauli string behaves approximately as if each of its constituents grows independently of each other (as long as the string size is much smaller than $n$, most sites in the string will not interact with each other at each layer). Therefore, in this case, the size grows as $\bar S_{t} = pe^{\lambda t}$, whereas the standard deviation grows as $\sqrt{p}e^{\lambda t}$. Choosing $p$ large enough ensures the peaked size mechanism at early times, where the above arguments are valid. For a fixed coupling constant $g$, using \eqref{FpureEst} with the above $\bar S$, we find that the first instance of perfect fidelity occurs at the time $t$ satisfying $(\frac{8}{5})^t = \frac{\pi n}{p g}$.

Let us turn to the three-boundary case. In addition to the analysis above, good fidelity requires that the operator distribution has little support on any of the $A_C$ qubits that are entangled with system $C$.

After a small number of time steps, the probability that a Pauli string $U E\sigma^{A_r}_{i,0}E^\dagger U^\dagger$ has support on a specific site is approximately $\bar S_t/(n+n_c)$. Therefore, the probability of having no support on the $A_C$ qubits, is $(1- \bar S_t/(n+n_c))^{n_c}$. Using $ \bar S_t = p e^{\lambda t}$, this probability can be close to one only at early times, when both $n_c$ and $p$ are much smaller than $n$. 
In this case, we obtain a time scale of
$\frac{1}{\lambda} \log{\frac{n}{p n_c}}$ for the decay of the fidelity.

In Fig. \ref{fig2:subfig_a}, we present the fidelity as a function of time with $n+n_c=10^5, \ p=100,$ and
$n_c=2$. The fidelity is seen in the orange curve, where good fidelity is demonstrated at early times. The blue curve describes the two-sided case, with the suppression at intermediate times, which happens due to large variance in the size distribution, as described in \cite{Schuster:2021uvg}. The green curve shows the upper bound in Eq.~\eqref{upperbound} on the fidelity represented by the orange curve. We can think of the green curve as isolating the effect that the sites $A_C$ have on the protocol, which ruins the perfect fidelity that would have been possible with a completely peaked size distribution and an optimal choice of $g$.

The above analysis demonstrates that the protocol is very sensitive to the dynamics. In $0D$, only a small number of $n_c$ sites and early times allow good fidelity. 

There is, however, one advantage in $0D$ in comparison to $1D$. Suppose that $A_C$ is very close to $A_r$, which is where the information is initially injected. In $1D$, the prospects for good fidelity are lost, as the information will soon reach $A_C$. In $0D$, however, there is no sense of geometric locality, and therefore good fidelity can still be achieved (on average) at intermediate times.

Another peculiar property of the $0D$ teleportation protocol is that $A_C$ can be part of the $p$ sites on which the initial information is encoded. 
With $1D$ dynamics, the probability that $U E\sigma^{A_r}_{i,0}E^\dagger U^\dagger$ has support on a site within the $p$ sites starts from $1$ at $t=0$ and saturates to $3/4$ after a short time. 
In $0D$ instead, if $p,n_c \ll n$ and at early times, at each time step, each site inside $A_C$ will most likely be coupled to a different site on which the operator $U E\sigma^{A_r}_{i,0}E^\dagger U^\dagger$ has no support.  A single site in $A_C$ that was within the support of the operator $U E\sigma^{A_r}_{i,0}E^\dagger U^\dagger$ before a given time step will remain within the support with probability $4/5$.  
By contrast, if the operator acts trivially on a site in $A_C$, that site will typically remain outside its support during the next time step, since it is likely to be paired with another site on which the operator also had no support.
Therefore, at early times, the average probability for support of the operator $U E\sigma^{A_r}_{i,0}E^\dagger U^\dagger$ at time $t$ on a given site on which it had support at time $t=0$ is roughly $(\frac{4}{5})^t$. At early times, the probability of losing support on all the $A_C$ sites if they were all initially with support is $(1-(4/5)^t)^{n_c}$. 
Requiring this probability to be at least $1-\epsilon$ gives
$
    t
    \gtrsim
    \frac{\log(n_c/\epsilon)}{\log(5/4)}.
$
Thus, for a fixed target probability, the relevant time scale grows
logarithmically with $n_c$.
This time should be before the 
time scale for which the probability of support in $A_C$ starts rising again due to the growth of the operator. The latter is given by the solution of $(4/5)^t = p e^{\lambda t}/n$, and describes when the operator is smeared with equal probability over all sites of the system, with no memory of the original $p$ sites.

In Fig. \ref{fig2:subfig_b}, the orange curve describes the fidelity as a function of time, with a single qubit in $A_C$ located within the $p$ initial sites. The effect of the $A_C$ sites can be seen by the green curve, which upper-bounds the orange curve, and limits both the early-time and late-time capabilities compared to the two-sided protocol. 
Following the discussion in the previous paragraph, the green curve, which is given by Eq.~\eqref{upperbound}, behaves as $1/4+3/4(1-(4/5)^t)$ roughly until the time that solves $(4/5)^t=p e^{\lambda t}/n$, after which the curve starts decreasing to its late-time value. For $\lambda=\log (8/5)$, $p=100$, $n\approx10^5$, this time is approximately $t\approx 10$ and matches the one observed in the figure.

\section{Generalization to qudits}
\label{sec:qudit_generalization}

In this section we discuss two distinct generalizations of the protocol. First, one may
increase the dimension of the teleported message. A qudit message can be treated by
replacing the EPR pair and Pauli operators by their
$d_{\mathrm{M}}$-dimensional counterparts. Similarly, several qubits
can be teleported in parallel. 

As discussed in
Ref.~\cite{Schuster:2021uvg}, teleportation of a multiple-qubit message in the two-boundary case is not possible at late times, see the argument around their equation (51). 
However, such a message can be teleported at intermediate times if the different qubits are inserted at well-separated sites in $1D$, or encoded to non-overlapping sites in $0D$. In this case, at intermediate times, the evolved
operators associated with the different input qubits have
approximately disjoint support. Their sizes are then additive, and the phase generated by the coupling factorizes. The problem, therefore, approximately decouples into independent single-qubit
teleportation channels. In the present setup, 
the same factorization allows good fidelity at intermediate times, but with the additional requirement that the individual operators do not reach the sites entangled with the third party.

A separate generalization is to 
take the resource systems to consist of qudits with local dimension
$q$, while keeping the teleported message to a single qubit (with local dimension $d_{\mathrm{M}}=2$). For simplicity, we will assume that $q$ is even. In this case, 
the insertion of the qubit into system $A$, \ie the SWAP operation in Eq.~\eqref{finalstate}, can be performed on a qubit subspace of $A_r$. For a fully scrambled operator on $L$ sites, each site carries a
nonidentity generalized Pauli operator with probability
\begin{equation}
    p_q=\frac{q^2-1}{q^2}.
\end{equation}
The mean and variance of the operator size are therefore 
\begin{equation}
    \frac{\bar{S}}{L}
    \simeq
    \frac{q^2-1}{q^2},
    \qquad
    \operatorname{Var}[S]
    \simeq L p_q(1-p_q)=
    L\frac{q^2-1}{q^4}.
\end{equation}
Increasing $q$ thus increases the typical fraction of sites on which the operator acts and decreases its relative size uncertainty. It also
changes the butterfly velocity. 

For one-dimensional (1D) Haar-random
brickwork circuits \cite{nahum2018operator},
\begin{equation}
    v_b(q)=\frac{q^2-1}{q^2+1}.
\end{equation}
At intermediate times, before the operator reaches the region
entangled with $C$, it has spread over approximately $2v_b(q)t$ sites,
so that
\begin{equation}
    \bar{S}_{AB}(t)
    \simeq
    2v_b(q)t\frac{q^2-1}{q^2}.
\end{equation}
The $q=2$ case was used to derive Eq.~\eqref{FpureApprox} from Eq.~\eqref{FLowerB}. For general $q$, the coupling required to generate the aligned teleportation phase is then 
\begin{equation}
    g_{\mathrm{opt}}(t;q)
    \simeq
    \frac{\pi n q^2}
    {2v_b(q)t(q^2-1)}.
\end{equation}
Taking the latest available time $t_c\simeq \Delta/v_b(q)$, when the operator front reaches system $C$, gives
\begin{equation}
    g_{\min}(q)
    \simeq
    \frac{\pi n q^2}
    {2\Delta(q^2-1)}.
\end{equation}
Changing the local dimension $q$ affects the 'width' of the light cone front of the evolving operator,
which scales like  $2 q\sqrt{t}/(q^2+1)$ \cite{nahum2018operator}.  This is the region in which the probability of support drops from $p_q$ to zero. 
This width fixes the temporal width of the decay of the fidelity, as was the case for $q=2$, see  Fig.~\ref{fig:subfig_a}.

In $0D$ circuit architectures, on the other hand, the Lyapunov exponent is $\lambda = \log(2q^2/(1+q^2))$~\cite{Schuster:2021uvg}. In addition, the average probability for support of the operator $U E\sigma^{A_r}_{i,0}E^\dagger U^\dagger$ at time $t$ on a given site on which it had support at time $t=0$ generalizes to approximately $(\frac{q^2}{q^2+1})^t$, which increases with $q$. This probability has been used in Sec.~\ref{sec:radncir0d} to explain the behavior of the green curve in Fig.~\ref{fig2:subfig_b}, which rises until the time scale at which the probability of support is equal on all the sites in $A$. This time scale is given by $\log_2(n/p)$, which is the solution of $(q^2/(q^2+1))^t = pe^{\lambda t}/n$. Increasing $q$ would decrease the rate at which the green curve, which upper-bounds the fidelity, rises until it reaches its maximum. Consequently, for a fixed $n$ and $p$, the maximal possible fidelity drops with $q$, while an increase of $n$ can compensate for this effect.

The local dimension therefore changes the relevant velocities,
operator-size statistics, and optimal coupling, but not the basic
three-boundary teleportation mechanism.
We leave a detailed numerical investigation of this generalization for future work.

\section{Conclusions}\label{sec:concl}

In this paper, we studied a three-party generalization of the many-body quantum teleportation protocol \cite{Schuster:2021uvg, brown2023quantum, nezami2023quantum}. We considered the situation in which three parties, $A, B$, and $C$, initially share a tripartite generalization of the TFD state, as defined in \cite{balasubramanian2014multiboundary, marolf2015hot, zou2022multiboundary}, and $A$ wishes to teleport a quantum state to $B$. In particular, we focused on the infinite-temperature limit, where the state initially consists of Bell pairs distributed between the three parties and contains only bipartite entanglement.

We derived a lower bound on the teleportation fidelity, see Eq.~\eqref{FLowerB}.
Unlike the previously studied two-sided protocol~\cite{Schuster:2021uvg, brown2023quantum, nezami2023quantum}, the fidelity here drops sharply once the information inserted in $A$ spreads to the sites that are entangled with the third party. The fidelity in both the original two-party setup and the present three-party generalization depends on the many-body dynamics of the systems. We focused on the $1D$ case under general assumptions of ballistic operator growth in local systems, with random brickwork circuits as a concrete example, and on the $0D$ case of fast-scrambling 2-local random circuits. To test the resulting predictions, we considered circuits composed of independently sampled Haar-random two-qubit gates. Their ensemble-averaged teleportation fidelities were evaluated numerically using random Clifford circuits with the same architecture, which reproduce the relevant Haar moments.

In $1D$, the picture is simple. The fidelity is influenced by two competing effects. Since the success of the protocol is based on an efficient scrambling of the message, the fidelity improves when the message is inserted at earlier times, as the operator has more time to scramble through the system. However, the presence of a third region starts to interfere at some point. The ballistic growth of operators sets a characteristic time $t_c$
at which the operator front reaches the region entangled with $C$, after which the fidelity quickly decreases. This time scale is approximately linear in the distance between the injection site and the region entangled with $C$. 
Our numerical results confirm this picture. Increasing the separation $\Delta$ from the region entangled with $C$ delays the fidelity drop and allows the maximal fidelity to approach unity. Our results are consistent with the predicted scalings for the critical time $t_c\sim\Delta/v_b$ and the required coupling  $g_{\min}\sim n/\Delta$.

In $0D$, the protocol succeeds only if the number of sites in $A$ entangled with $C$ is much smaller than the number of sites in $A$ entangled with $B$. Unlike in $1D$, the protocol can also succeed when the information is inserted on sites entangled with $C$ since there is some probability of losing support on those sites under time evolution. The main phenomenological difference in $0D$ between the two-sided and three-sided protocols is the reduced fidelity at late times.
For insertions of the message on the $A_C$ sites, the fidelity of the three-sided protocol is also limited at early times.
Our numerical results confirm both effects: when the message is inserted outside $A_C$, a high-fidelity window is followed by the characteristic late-time suppression, whereas insertion in $A_C$ can produce an intermediate-time fidelity peak as the dynamics first remove and later regenerate operator support on those sites.

Several interesting observations arise from the results. First, whereas the two-sided protocol is sensitive to the operator-size distribution, the presence of a third party makes the fidelity
additionally sensitive to whether the evolved operator has support on
the degrees of freedom entangled with that party. In this sense, the
three-party protocol provides a subsystem-resolved
probe of operator spreading.

Second, in $1D$, the minimal coupling required for high fidelity scales inversely with the distance between the injection site and the region entangled with $C$. An analogous spatial dependence appears in hot holographic multi-boundary wormholes: a signal can reach a chosen asymptotic boundary only when injected from an appropriate angular region in which the state contains enough bipartite entanglement between the two boundaries. In holography, the null shift generated by the double-trace deformation involved in the teleportation protocol must exceed the gap produced by the causal shadow, implying an angle-dependent threshold for traversability \cite{al2021traversability}. The causal-shadow geometry underlying this construction was studied in Refs.~\cite{balasubramanian2014multiboundary, marolf2015hot}.

One caveat, however, is that in the infinite-temperature limit, the
distance between the horizons across the causal-shadow region is
expected to shrink to zero, reflecting the purely bipartite
entanglement of the initial state \cite{peach2017tensor}.
Nevertheless, for teleportation from $A$ to $B$, the degrees of
freedom in $A_C$ provide an operational analogue of a causal shadow:
once the message reaches them, recovery in $B$ is suppressed. 
The
distance to the $A_C$ sites depends on the message insertion point, producing spatial dependence of both the characteristic time for the protocol's failure and the coupling required for successful teleportation.
In holography, a finite causal shadow is instead associated with
multipartite entanglement. Although such entanglement is absent from
our initial state, generic scrambling is expected to generate it
dynamically once the evolved message operator has support on both
$A_B$ and $A_C$, which happens approximately at the same time at which the teleportation fidelity
begins to decrease. Although our models are far from holographic, the
microscopic origin of the coupling and insertion location dependence within our protocol may
still be relevant for holographic systems.

Third, the intermediate-time peak in the three-sided $0D$ fidelity in Fig.~\ref{fig2:subfig_b} is qualitatively similar in shape to the peak near the scrambling time in two-sided holographic teleportation at finite temperature \cite{brown2023quantum,Schuster:2021uvg}. A potential explanation for this similarity can be drawn from tensor-network and quasiparticle-inspired pictures of the thermofield double. A local boundary insertion initially excites near-boundary degrees of freedom, while deeper degrees of freedom in the bulk heuristically carry more of the entanglement between the two sides. These roles are loosely analogous to those of $A_C$ and $A_B$ in our three-system protocol, respectively. In Fig.~\ref{fig2:subfig_b}, the fidelity becomes large as the message loses support on $A_C$, and decreases when that support grows again. This comparison is qualitative and is distinct from the size-winding mechanism, usually used to explain holographic teleportation. Making it precise would require more work, which we leave for the future.

Fourth, although our analysis has focused on information transfer
between two of the three parties, the setup can be modified to form
a three-node network in which the injection point selects the
receiver. If the coupling between $A_B$ and $B_A$ is supplemented by
the analogous coupling between $A_C$ and $C_A$, then, for suitable
$g$ and $t$, a message inserted in $A_B$ is recovered in $B$, whereas
a message inserted in $A_C$ is recovered in $C$. Thus, the
construction provides a destination-selective three-node
teleportation protocol.

The above discussion leads to several future directions. The first is to test the protocol in microscopic systems that exhibit holographic features in $d>0$. One possible example is the $(1+1)$-dimensional generalization of the SYK model of \cite{Altland:2025qqw}.

A further direction is to test the protocol at finite temperatures and study how it depends on quantum informational quantities of the initial state that have a holographic interpretation. One such quantity is the mutual information between subregions in $A$ and $B$, which is related to the lengths of geodesics connecting the edges of the subregions \cite{ryu2006holographic}. 
Suppose that we take in $A$, which has $2n$ sites, a subregion of size $n$ centered
at the insertion point $r$ (measured with respect to the $A_B-A_C$ interface),  together with the corresponding
subregion in $B$. In the infinite-temperature initial state we considered, the difference between
this mutual information and its value when $r$ lies at the
$A_B-A_C$ interface is exactly $2\Delta$ bits, where $\Delta$ is
defined in Eq.~\eqref{eq:Delta}.
Consequently, the scaling $g_{\min}\sim n/\Delta$ may equivalently
be expressed as an inverse dependence on this excess mutual
information. It would be worthwhile to test whether an analogous
relation persists at finite temperature. 

Introducing finite temperatures could also help reveal the role of multipartite entanglement in the protocol's success.
One challenge at finite temperature is that the initial state and the
real-time evolution must be constructed from the same \emph{time-independent} Hamiltonian, so
the random-Clifford ensemble used here is no longer directly
applicable. Exact numerical simulations are then restricted to small
systems, although quantum simulators may provide access to larger
system sizes.

Another direction is to extend the present spatially-resolved picture
to more general ways of resolving operator spreading. In the
construction studied here, the pattern of entanglement with $B$ and
$C$ singles out the factors $\mathcal{H}_{A_B}$ and
$\mathcal{H}_{A_C}$, making the teleportation fidelity sensitive to
the operator weight associated with each factor. More generally,
suitably chosen resource states and inter-system couplings might
probe operator weight in selected tensor factors, symmetry sectors,
or subalgebras of observables. It would be useful to determine which
such structures admit a teleportation-based diagnostic and how the
resulting fidelities encode the corresponding operator-weight
distributions.

Beyond the three-party setting, it would be valuable to extend the setup to a
network of many parties and study information transfer across it.
Inspired by Ref.~\cite{may2021interpolating}, one may ask whether
tuning the entanglement pattern and the coupling between the different nodes can
produce boundary information-transfer dynamics with a dual
description as propagation through geometries interpolating between
multi-boundary wormholes and connected bulk geometries. 
In such a case, the transfer between nodes might be described by causal bulk trajectories through the resulting connected geometry, with the trajectories depending on the insertion profile, in the spirit of Ref.~\cite{bamba2024spacetime}. These connections should be regarded as motivations for future work rather than direct consequences of the present study.

Finally, it would be interesting to implement the protocol on quantum
processors or simulators. Such experiments could test the
subsystem-resolved scrambling diagnostic introduced here, including
its robustness to noise, while also providing access to
finite-temperature and genuinely multipartite resource states that
are difficult to study with the numerical methods used in this work.

\section*{Acknowledgments}
We would like to thank Dvir Cohen, Khen Cohen, Yaron Oz and Michael Walter for useful discussions.
The work of S.C. is supported by the Israel Science Foundation (grant No. 1417/21), by the German Research
Foundation through a German-Israeli Project Cooperation (DIP) grant “Holography and
the Swampland”, by Carole and Marcus Weinstein through the BGU Presidential Faculty
Recruitment Fund, by the ISF Center of Excellence for theoretical high energy physics, by the VATAT Research Hub in the Field of quantum computing and by the ERC starting Grant dSHologQI (project number 101117338). T.S. acknowledges support for this work from ITAMP, funded
by the US National Science Foundation, and from the VATAT Outstanding Postdoctoral Fellowship in Quantum Science and Technology. T.S. is grateful for the generous support from the RH Growth Foundation.

\bibliographystyle{unsrt}
\bibliography{bibliography}

\end{document}